\RequirePackage{lineno}
\documentclass[preprint,a4paper,nofootinbib]{revtex4-1}
\usepackage[utf8]{inputenc}
\usepackage{textcomp}
\usepackage{gensymb}
\usepackage{amsmath}
\usepackage{amssymb}
\usepackage{xspace}
\usepackage{nicefrac}
\usepackage{xcolor}
\usepackage{graphicx}
\usepackage{soul}
\usepackage{hyperref}
\usepackage[caption=false]{subfig}
\usepackage{multirow}
\usepackage{cleveref}
\usepackage[titletoc,toc,page]{appendix}

\begin{document}

\title{Magnetically-induced anisotropies in the arrival directions of ultra-high energy cosmic rays from nearby radio galaxies}

\date{\today}
\author{Cainã de Oliveira}
\email{caina.oliveira@usp.br}
\author{Vitor de Souza}
\email{vitor@ifsc.usp.br}
\affiliation{Instituto de F\'isica de S\~ao Carlos, Universidade de S\~ao Paulo, Av. Trabalhador S\~ao-carlense 400, S\~ao Carlos, Brasil.}

\begin{abstract}

Detailed simulations of the arrival directions of ultra-high energy cosmic rays are performed under the assumption of strong and structured extragalactic magnetic field (EGMF) models. Particles leaving Centaurus A, Virgo A, and Fornax A are propagated to Earth, and the simulated anisotropic signal is compared to the dipole and hotspots published by the Pierre Auger and Telescope Array Collaborations. The dominance of the EGMF structure on the arrival directions of events generated in local sources is shown. The absence of events from the Virgo A direction is related to the strong deviation caused by the EGMF. Evidence that these three sources contribute to an excess of events in the direction of the three detected hotspots is presented. Under the EGMF considered here, M82 is shown to have no contribution to the hotspot measured by the Telescope Array Observatory.

\end{abstract}

\maketitle

\section{Introduction}

The detection of ultra-high energy cosmic ray sources is one of the most important open questions in astrophysics. The identification of only one source would open a new window of discoveries concerning the balance and transfers of energy in extremely powerful objects. The lack of knowledge about the source position and type introduces several uncertainties in the interpretation of the data, representing a major barrier for the progress towards describing the most energetic phenomena in nature. Given that ultra-high energy cosmic rays (UHECR) are charged, the main challenges to overcome are the unknowns about the structure and intensity of the magnetic fields in the Universe.

The Pierre Auger~\citep{2015172} and the Telescope Array~\citep{KAWAI2008221} Observatories have detected the more precise UHECR signals from the sky. Three regions with relative excess of events (hotspots) in the arrival directions of UHECR have been identified, two regions in the Pierre Auger Observatory data~\citep{aab2018indication} and one region in the Telescope Array data~\citep{Abbasi_2014}. A small level of anisotropy was also detected in the arrival directions of events measured by the Pierre Auger Observatory, in which a dipole signal was reported~\citep{1266,Aab_2018}. A correlation of the arrival directions of events with the supergalactic plane has been reported by the Telescope Array Collaboration~\citep{Abbasi_2020} and an indication of a correlation with starburst regions has been reported by the Pierre Auger Observatory~\citep{aab2018indication}.

The relation of these signals with the sources is not straightforward because deviations caused by the magnetic fields on the way to Earth blur, relocate or erase the direction of the source in the sky maps. The importance of the Galactic and extragalactic magnetic field intensity and structure has been explored in the past by many authors~\citep{SIGL2004224,Tanco2001,Medina_Tanco_1998,1995ApJ...455L..21L,Jansson_2012,lang2020revisiting,ERDMANN201654}. The dipole measured by the Pierre Auger Observatory points away from the Galactic plane and center suggesting an extragalactic origin of UHECR. The amplitude of the dipole increase with energy above 4 EeV suggesting an increase in the relative contribution to the flux arriving at Earth from the nearby sources~\citep{1266,Aab_2018}. Despite the exact coordinates of one of the hotspots measured by the Pierre Auger Observatory was not published, it seems to be related to the direction of the radio galaxy Centaurus A~\citep{aab2018indication}. The starburst region M82 has been proposed as the source of the hotspot measured by the Telescope Array Observatory~\citep{Abbasi_2014}. The lack of an excess in the data of the Pierre Auger Observatory in the direction of Virgo A has called the attention of several authors~\citep{10.1093/mnras/stz094,Dolag_2009}.

Radio galaxies have been considered prime source candidates~\citep{ginzburg1963cosmic}. Several models have been proposed to describe acceleration mechanisms able to generate UHECRs in radio galaxies. Most of the models correlate the UHECR flux emitted by radio galaxies with the jet power~\citep{pacholczyk1970radio,rachen1993extragalactic,Eichmann_2018,eichmann2019high,matthews2019ultrahigh}, which in turn can be related to the radio luminosity~\citep{godfrey2016mutual,cavagnolo2010}. In this scenario, the radio galaxies Centaurus A (NGC 5128), Virgo A (NGC 4486), and Fornax A (NGC 1316) should produce a significant percentage of the UHECR flux arriving at Earth because they have a radio flux about one order of magnitude above other nearby radio galaxies~\citep{ginzburg1963cosmic, van2012radio} and because of their proximity ($< 21$ Mpc). Besides that, it has been pointed out that nearby sources are needed to explain not only the arrival direction data, but also the energy spectrum and composition measurements~\citep{taylor2011need,lang2020revisiting,PhysRevD.103.063005}.

In this paper, detailed simulations of arrival direction maps of particles produced in Centaurus A, Virgo A, and Fornax A are presented. The details about the simulations are given in section~\ref{sec:method}. A special treatment of the local extragalactic magnetic field is considered. In comparison to the methodology of previous works~\citep{SIGL2004224,Tanco2001,Medina_Tanco_1998,1995ApJ...455L..21L,PhysRevD.68.043002}, this paper introduces recent models of the structure of the local extragalactic magnetic field~\citep{hackstein2018simulations} in full three-dimensional simulations~\citep{batista2016crpropa}. It is also shown here for the first time that the structure of the local extragalactic magnetic field is determinant to understand the signals of UHECR anisotropy detected by the Pierre Auger and Telescope Array Observatories.

The effects of the structure of the extragalactic magnetic fields are discussed in section~\ref{sec:arrival}, and the deficit in the direction of Virgo A is explained. The dipole signal is explored in section~\ref{sec:dipole}, and the possibility to explain its direction only with events from Centaurus A, Virgo A, and Fornax A is discussed. The hotspots are discussed in section~\ref{sec:hotspots} and possible counterparts with Centaurus A, Virgo A, and Fornax A are presented. The work is summarized in section~\ref{sec:conclusion}.

\section{Method: the sources, the medium and the propagation}
\label{sec:method}

Some source types have been proposed to accelerate UHECR~\citep{doi:10.1146/annurev.aa.22.090184.002233,OLINTO2000329,Wang_2008,Dermer_2010,10.1111/j.1745-3933.2008.00547.x,Halzen_2002,PhysRevD.58.123005,1995ApJ...453..883V,PhysRevLett.75.386,1989A&A...221..211M}, among which radio galaxies are one of the most accepted candidates~\citep{matthews2019ultrahigh,bell2018cosmic,Eichmann_2018,rachen2008ultrahigh,ROMERO1996279,1995ApJ...454...60N}. Centaurus A (Cen A), Virgo A (Vir A), and Fornax A (For A) are the most powerful nearby radio galaxies. Their radio luminosity at frequency 1.1 GHz, $L_{1.1}$, are $L_{1.1}^{\rm CenA} = 2.6 \times 10^{40}$ erg/s, $L_{1.1}^{\rm VirA} = 7.6 \times 10^{40}$ erg/s, and $L_{1.1}^{\rm ForA} = 8.3 \times 10^{40}$ erg/s~\citep{van2012radio}, an order of magnitude higher than other nearby radio galaxies. They are relatively close to Earth with distances: $D^{\rm CenA} = 3.8$ Mpc, $D^{\rm VirA} = 18.4$ Mpc, and $D^{\rm ForA} = 20.9$ Mpc. The combination of high radio power and proximity makes them strong candidates in dominating the UHECR sky. Reviews about these sources can be found in references~\citep{cena00,refId0,Ferrarese_2006,Macchetto_1997,m87book,Jensen_2003,1980ApJ...237..303S,10.1111/j.1365-2966.2008.13960.x}.

In this paper, UHECRs are injected from Cen A, Vir A, and For A and propagated to Earth on computational simulations. Each of the three sources has been extensively studied as possible UHECR sources~\citep{ROMERO1996279,matthews2018fornax,10.1093/mnras/stz094,biermann2010ultra,biermann2012centaurus,wykes2013mass,kachelriess2009high}. It is not the scope of this paper to discuss the acceleration models in the sources. Therefore, a standard first-order Fermi acceleration~\citep{kotera2011astrophysics} in diffusive shock of AGN jets~\citep{matthews2019ultrahigh} has been used. In this scenario, the UHECR injection spectrum is given by a power-law with a charge-dependent exponential cutoff~\citep{supanitsky2013upper}. Details are given in appendix~\ref{app:emission}.

UHECRs interact with radiation and magnetic fields on the way from the source to Earth changing its properties. The CRPropa3 framework~\citep{batista2016crpropa} was used to propagate particles from their sources to Earth. A very time-consuming 3D simulation was done with $3\times10^8$ events for each source, for each nuclei: $^1$H (proton), $^2$He, $^7$N, $^{14}$Si and $^{26}$Fe, with injected energies between $10^{18}$ and $10^{21}$ eV following a power-law spectrum with spectrum index $-1$ to guarantee equal statistical uncertainties at all energies. The particles were emitted isotropically from the source. After propagation, the energy spectrum was weighted according to the explanation in appendix~\ref{app:emission}. The energy loss mechanisms and interactions considered were pion photoproduction, pair production, photo-disintegration, nuclear decay, and adiabatic loss. UHECR are propagated until they reached the surface of a sphere with radius ($r_{obs}$) centered on Earth. The radius ($r_{obs}$) was set to $r_{obs}^{\rm CenA} = 100$ kpc, $r_{obs}^{\rm VirA} = 400$ kpc and $r_{obs}^{\rm ForA} = 500$ kpc according to the source in consideration. The variable value of $r_{obs}$ minimizes the number of simulated events saving computational time and guarantees an uncertainty in the arrival direction of events from each source~\citep{10.1093/mnras/stw1903} smaller than the uncertainties in the arrival direction measurements~\citep{Aab_2020}. A smaller set of events was also simulated for an observational radius of 10 kpc for all sources. The conclusions presented here remain the same irrespective of $r_{obs}$. The artificial uncertainty due to the finite observer can be estimated by $\arcsin{(r_{obs}/D})$, if the magnetic field in the observer neighborhood is not to much intense, therefore if $r_{obs} = 100$ kpc and $D_{source} = 10$ Mpc then the artificial deflection is $0.01 < 1$ and if $r_{obs} = 10$ kpc the artificial deflection becomes one order of magnitude smaller.

The background photon fields considered in this paper were the cosmic microwave background and the extragalactic background light~\citep{gilmore2012semi}. The photon fields have an important effect on the propagation of UHECR even for nearby sources. The interaction of UHECR with the gas surrounding the sources can be safely neglected given the baryon content estimated in the filaments of massive galaxy clusters to have a density $\approx 10^{-5}$ cm$^{-3}$~\citep{density}, the mean free path for proton-proton interactions in this medium is of the order of 1 Gpc.

The Galactic magnetic field (GMF) is known to have a large effect on the arrival direction of UHECR~\citep{Farrar_2013,ERDMANN201654}. For all plots and analyses in this paper, the most updated GMF model of reference~\citep{Jansson_2012}, JF12, was used. The regular and random components were taken into account. This model includes the analysis of the WMAP7 Galactic Synchrotron Emission map and forty thousand extragalactic rotation measures. Since the Milky Way is significantly smaller than the closest source distance considered here ($D^{\rm CenA} = 3.8$ Mpc), UHECR energy losses inside our galaxy were neglected, allowing the net effect of the GMF in the arrival direction of UHECR to be accounted for using parametrizations~\citep{BRETZ2014110}.

The extragalactic magnetic field (EGMF) is also known to have a large effect on the arrival directions of UHECR~\citep{SIGL2004224}. The EGMF intensity can be inferred by several techniques including, gamma-ray energy spectrum modulations of distant sources, Faraday rotation in the polarized radio emission from distant sources, and properties of the cosmic microwave background radiation~\citep{Kronberg_1994,durrer,Subramanian_2016}. Because these are integrated measurements that depend mainly on the perpendicular component of the field, nothing can be extracted from these data about the EGMF topology~\citep{widrow_2012}. Given the uncertainties, simple cellular structure EGMF models~\citep{Tanco2001}, based on cosmological arguments of seed fields~\citep{durrer}, with $\sim 10^{-9}$ G intensity and $\sim$ Mpc correlation length, have been widely used in UHECR propagation studies~\citep{Waxman_1996,Aloisio_2004,Batista_2014}. A reference Cellular EGMF characterized by cells with 0.37 Mpc size in which the field has random orientation and intensity given by a Kolmogorov spectrum with root mean square 3 nG was used in some studies below for comparison.

Structured models of the local EGMF with void and filaments have been long proposed~\citep{Ryu909,Cho_2009}. It has been shown that the structure of the EGMF leads to large deflections ($>90^\circ$ below the GZK limit) in the arrival directions of particles generated in local sources~\citep{Medina_Tanco_1998,Das_2008}. In this paper, the most updated simulation of the local EGMF is used~\citep{hackstein2018simulations}. This reference calculates the local EGMF structure from two fundamental hypotheses named \textit{primordial} and \textit{astrophysical}. The \textit{primordial} models start with a non-negligible EGMF at $z=60$. The \textit{astrophysical} one was modeled as impulsive thermal and magnetic feedback in halos starting with a uniform lower magnetic field level of $10^{-20}$ G at $z= 60$. Magneto-hydrodynamical simulations evolve the EGMF to the present time taking into consideration several constraints in a $\Lambda$CDM Universe. In this paper, three EGMF models from reference~\citep{hackstein2018simulations} are used, covering the widest range of field intensity and level of structure:  AstrophysicalR (AstroR), Primordial2R (Prim2R) and  Primordial (Prim), where Prim starts with a uniform 0.1 nG intensity field, while Prim2R starts with a power-law distribution of spectral index -3 and $B_{rms} = 1$\ nG. Maps of the intensity of these fields with the position of the sources studied here are shown in references~\citep{hackstein2018simulations,deoliveira2021probing}.

\section{Results: Arrival directions of UHECR from Cen A, Vir A, and For A}
\label{sec:arrival}

\begin{figure}
  \centering
  \includegraphics[width=1.0\columnwidth]{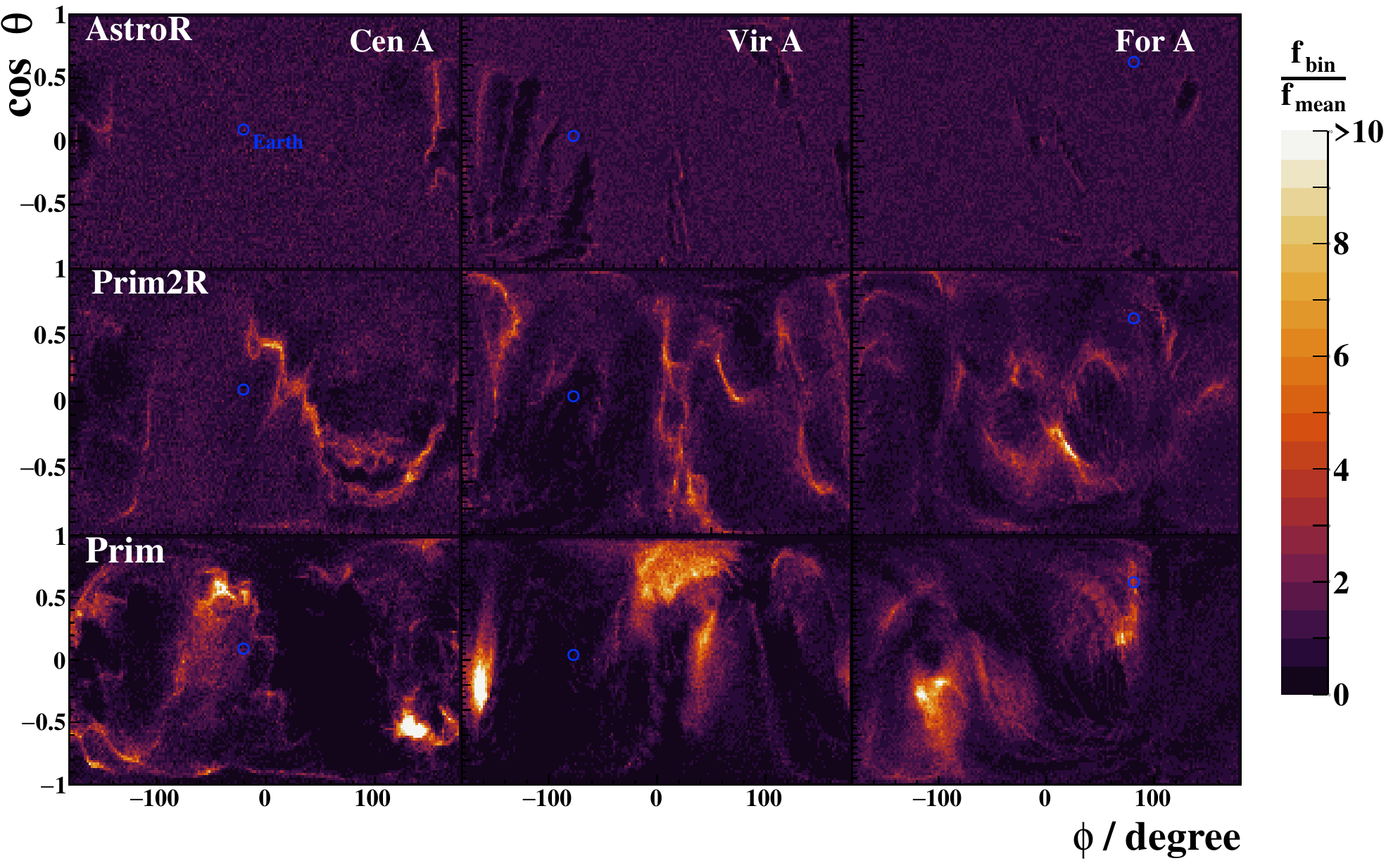}
  \caption{Relative flux of UHECR around Cen A, Vir A, and For A. Each line in the figure shows one EGMF model: AstroR, Prim2R, and Prim. Each column in the figure shows the results for one source: Cen A, Vir A, and For A. Particles were tracked from the source (center of the map) until they reached a sphere with radius equal to the distance from the source to Earth. The maps show the arrival position of all particles in this sphere. The blue circle shows the position of Earth. The color code in the maps represents the relative flux of UHECR. All nuclei are considered to leave the source with equal flux.}
  \label{fig:maps:sources:center}
\end{figure}

Figure~\ref{fig:maps:sources:center} shows the relative flux of UHECR in small portions of the sky around Cen A, Vir A, and For A. Each line in the figure shows one EGMF model: AstroR, Prim2R, and Prim. Each column in the figure shows the results for one source: Cen A, Vir A, and For A. Particles were tracked from the source (center of the map) until they reach a sphere with radius equal to the distance from the source to Earth. The maps show the arrival position of all particles in this sphere. The blue circle shows the position of Earth. The color code in the maps shows the relative flux of UHECR. All nuclei are considered to leave the source with equal flux. Similar maps, leading to the same conclusions, with only protons and only iron nuclei leaving the source are shown in appendix~\ref{app:nuclei}. There are clear regions with enhanced and suppressed flux for each source and EGMF model. The effect of the structure of the EGMF is very evident.

The relative flux of UHECR arriving at Earth from each source is strongly affected by the local structure of the EGMF. The flux at Earth position is not largely affected by the EGMF choice when Cen A is considered. Earth is in a flux-suppressed position to receive UHECR from Vir A. The EGMF structure diverts UHECR from Vir A, causing a lack of event from its direction. The reduction of events from Vir A is very clear in figure~\ref{fig:maps:sources:center} for the Prim2R and Prim models, and despite being less visible, it is also present for the AstroR model. Earth is in a favored position to receive UHECR from For A. The EGMF structure focuses UHECRs from For A in the direction of Earth, as can be clearly seen in the maps.

\begin{figure}
  \centering
  \includegraphics[trim=0 350 0 0,clip,width=1.0\columnwidth]{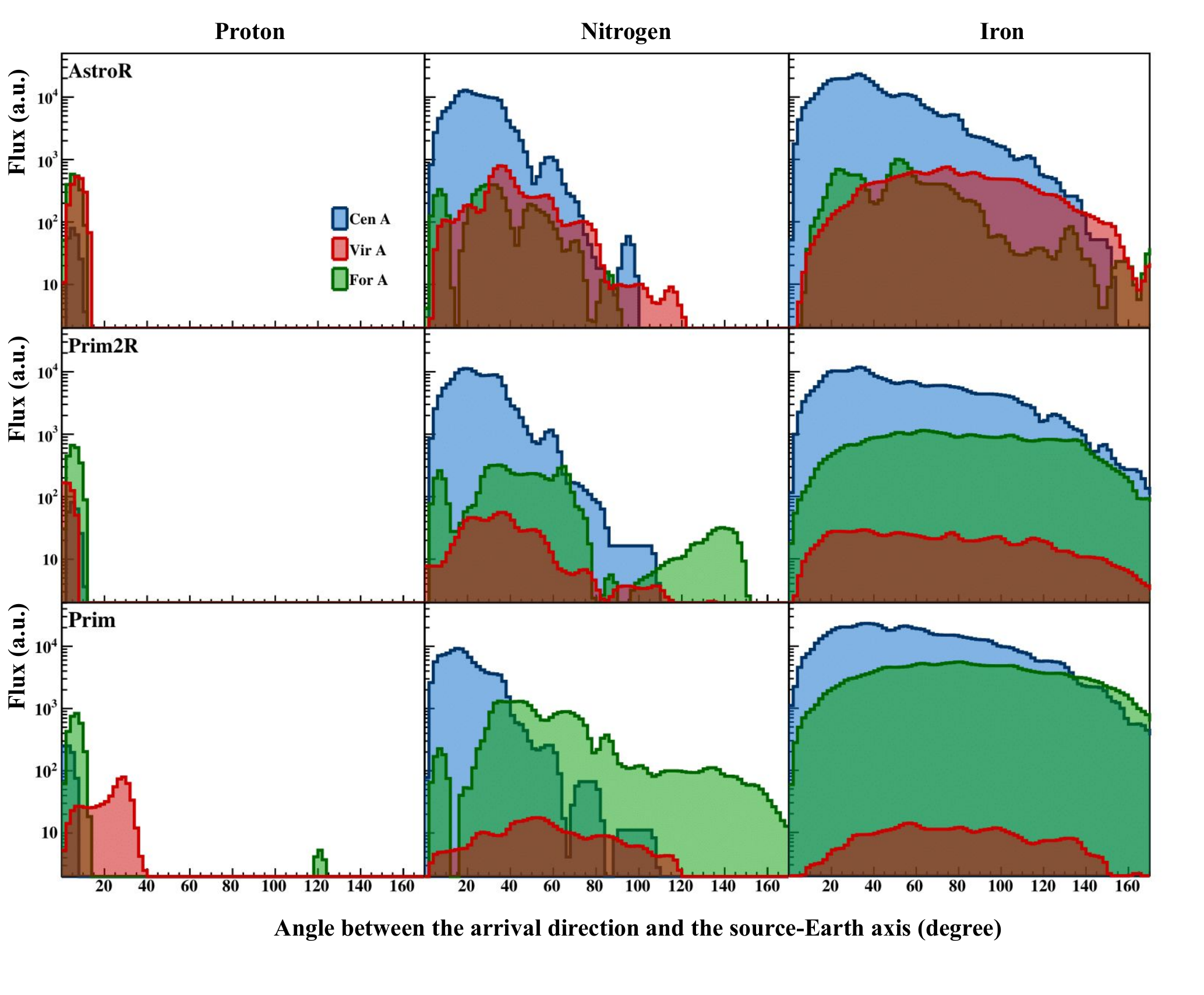}
  \caption{Angular distribution of events with $E > 32$~EeV in relation to the source direction. Each line in the figure shows one EGMF model: AstroR, Prim2R, and Prim. Each column in the figure shows a different nucleus leaving the source: proton, nitrogen, and iron nuclei. Note that in each column of the figure, all nuclei fragments on the way to Earth are shown as arriving at Earth when only proton, nitrogen, or iron nuclei left the source. The three sources are differentiated by colors: blue, red, and green for Cen A, Vir A, and For A, respectively. The sources are considered to output the same UHECR flux.}
  \label{fig:deflection:hist}
\end{figure}

Figure~\ref{fig:deflection:hist} shows the angular distribution of events in relation to the source direction. Each line in the figure shows one EGMF model: AstroR, Prim2R, and Prim. Each column in the figure shows a different nucleus leaving the source: proton, nitrogen, and iron nuclei, representing light, intermediate, and heavy composition leaving the sources. Similar plots, leading to the same conclusions for helium and silicon nuclei, are shown in appendix~\ref{app:nuclei}. Note that in each column of the figure, all nuclei fragments on the way to Earth are shown as arriving at Earth when only proton, nitrogen, or iron nuclei left the source. The three sources are differentiated by colors: blue, red, and green for Cen A, Vir A, and For A, respectively. The sources are considered to output the same UHECR flux. Events are deviated by large angles ($\sim20^\circ$) from the source direction even for the nearest source (Cen A). The average deflection depends on the charge of the nuclei leaving the source, being smaller for protons and resulting in an almost uniform distribution for iron nuclei.

The structure of the EGMF interferes clearly with the flux received from each source. For a relatively weak and less structured field (AstroR), the relative number of events arriving at Earth from each source is to a good approximation determined only by the geometric effect. However, the field structure becomes dominant for a stronger and structured EGMF (Prim and PrimR). Note in figure~\ref{fig:deflection:hist} the larger number of events arriving at Earth from For A than from Vir A when iron nuclei are simulated (right column). If only the geometric effect given by distance under isotropic emission is considered, the flux arriving at Earth from For A would be 1.3 times smaller than the flux from Vir A. When the effect of the EGMF Prim2R (Prim) is taken into account, the flux arriving at Earth from For A is 38 (443) times larger than the flux from Vir A.

\subsection{The dipole}
\label{sec:dipole}

The Pierre Auger Collaboration published a multipole analysis of the arrival directons of detected events in which a dipole signal was detected~\citep{1266,Aab_2018,deAlmeida:20212Z}. In this analysis, the data set was divided into five energy bins: $\geq 8$, $4-8$, $8-16$, $16-32$ and $\geq 32$ EeV. For each energy bin, a dipole was detected at a given direction and amplitude (see table 1 of reference ~\citep{deAlmeida:20212Z}). The p-values under the null hypothesis of isotropy for the first-harmonic modulation in right ascension are $5.1\times10^{-11}$, 0.14, $3.1\times10^{-7}$, $7.5\times10^{-4}$, 0.01 respectively for the energy bins.

Recent works have shown the importance of the local sources in the determination of the dipole direction~\citep{PhysRevD.103.063005}. In this section, the dipole directions for the same energy bins considered in the analysis done by the Pierre Auger Collaboration are calculated with the events arriving at Earth from Cen A, Vir A, and For A according to the simulations explained above. The dipole directions generated by the arrival directions of UHECR simulated from these three sources is compared with the measurement done by the Pierre Auger Observatory. The calculation of the dipole direction with only three sources is a simplification because further sources certainly influence the dipole direction as well. The idea is to test how dominant is the influence of these three sources in the direction of the dipole by ignoring the contribution of other sources. It is expected that nearby sources contribute more significantly to the flux at the highest energies because of the GZK suppression~\citep{Greisen1966,bib:zk}.

\begin{figure}
  \centering
  \includegraphics[angle=270,width=1.0\columnwidth]{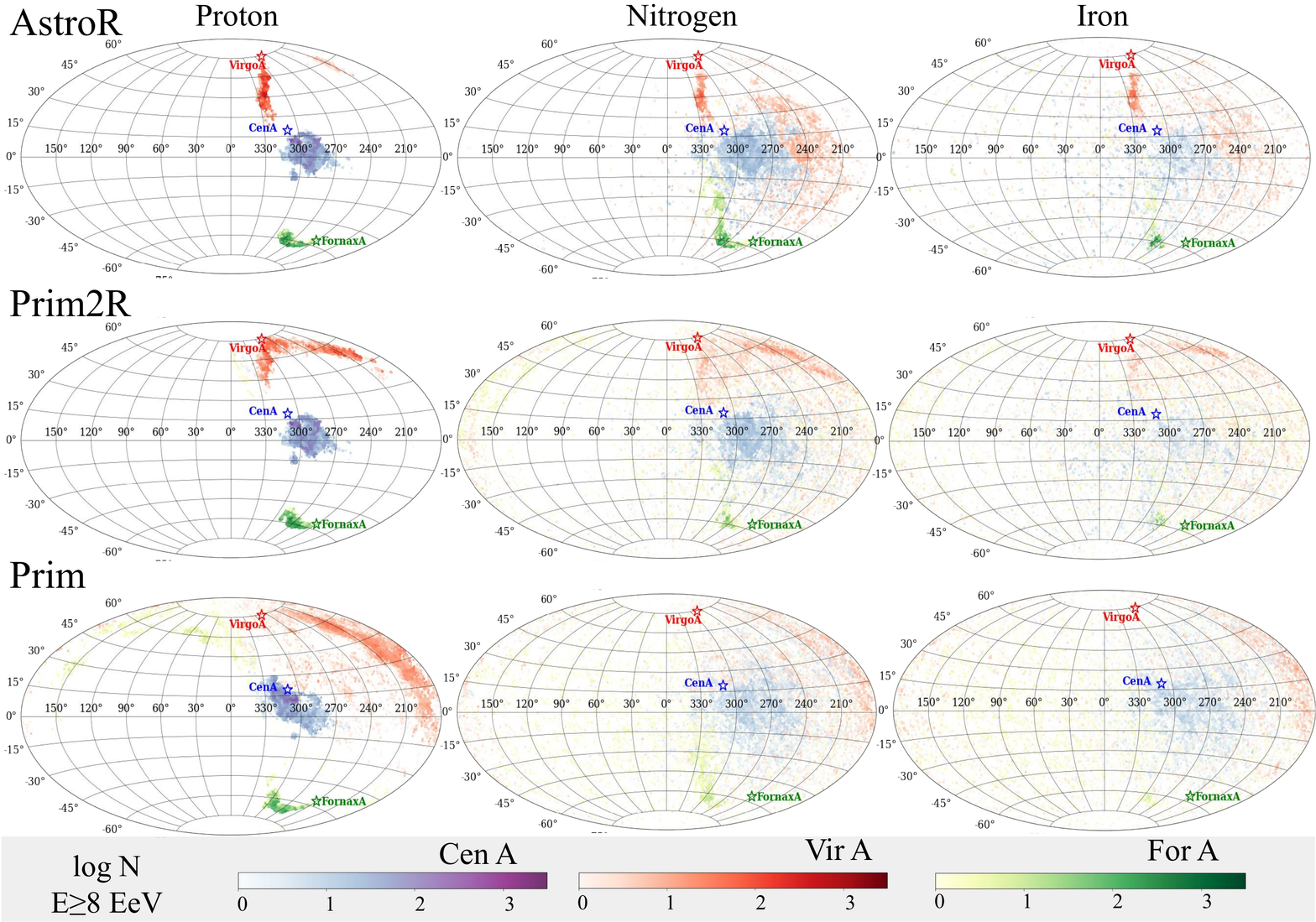}
  \caption{Sky maps in Galactic coordinates and Aitoff projections of all simulated events which arrived at Earth with energy above 8 EeV. Each line in the figure shows one of the EGMF models considered here. Each column in the figure shows a different nucleus leaving the source: proton, nitrogen and iron nuclei. Note that in each column of the  figure, all nuclei fragments on the way to Earth are shown as arriving at Earth when only proton, nitrogen, or iron nuclei left the source. The three sources are shown as blue, red, and green stars for Cen A, Vir A, and For A, respectively. The flux of events follows the same color-code, each color representing only the events generated in the respective source.}
  \label{fig:arrival:8:pr_n_fe}
\end{figure}

\begin{figure}
  \centering
  \includegraphics[angle=270,width=1.0\columnwidth]{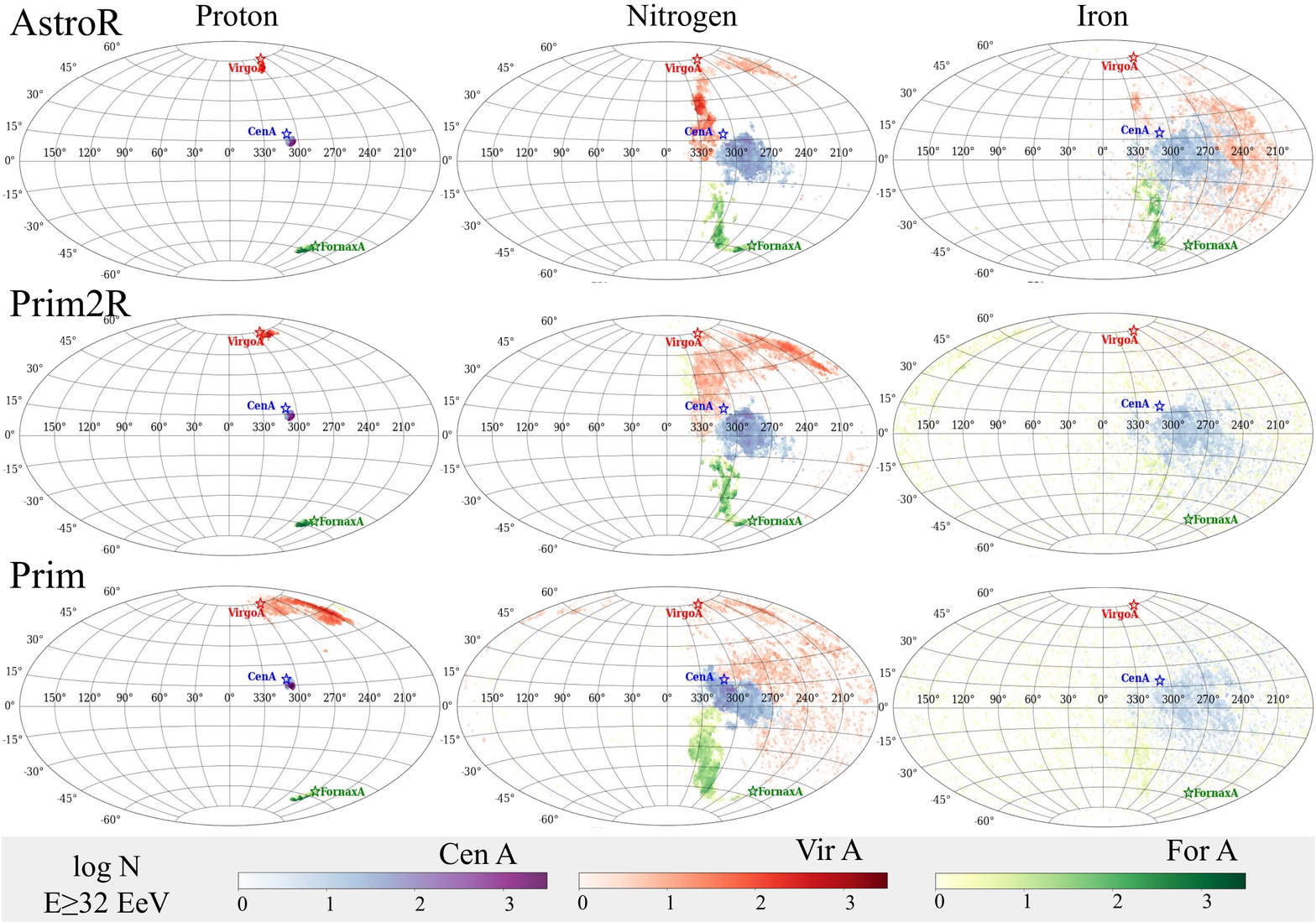}
  \caption{Sky maps in Galactic coordinates and Aitoff projections of all simulated events which arrived at Earth with energy above 32 EeV. Each line in the figure shows one of the EGMF models considered here. Each column in the figure shows a different nucleus leaving the source: proton, nitrogen, and iron nuclei. Note that in each column of the figure, all nuclei fragments on the way to Earth are shown as arriving at Earth when only proton, nitrogen, or iron nuclei left the source. The three sources are shown as blue, red, and green stars for Cen A, Vir A, and For A, respectively. The flux of events follows the same color-code, each color representing only the events generated in the respective source.}
  \label{fig:arrival:32:pr_n_fe}
\end{figure}

Figures~\ref{fig:arrival:8:pr_n_fe} and~\ref{fig:arrival:32:pr_n_fe} show the arrival directions of the simulated events which arrived at Earth with energies above 8 and 32 EeV, respectively. Each line in the figures refers one of the EGMF models considered here. Each column in the figures shows a different nucleus leaving the source: proton, nitrogen, and iron nuclei. Similar plots, leading to the same conclusions for helium and silicon nuclei, are shown in appendix~\ref{app:nuclei}. Note that in each column of the figure, all nuclei fragments on the way to Earth are shown as arriving at Earth when only proton, nitrogen, or iron nuclei left the source. The three sources are shown as blue, red, and green stars for Cen A, Vir A, and For A, respectively. The flux of events follows the same color-code, each color representing only the events generated in the respective source. The importance of the structure of the local EGMF is evident in many ways.

The overall conclusions in figures~\ref{fig:arrival:8:pr_n_fe} and~\ref{fig:arrival:32:pr_n_fe} are the same for the two energy ranges above 8 and 32 EeV with the expected caveat that lower energy events are more spread. Protons leaving all three sources (first column) suffer a negligible deflection in the AstroR and Prim2R EGMF models. However, protons leaving Vir A are deviated in the Prim EGMF model by large angles, thus arriving at Earth with directions pointing several degrees away from Vir A. Note the large number of events along longitude 210$^\circ$ crossing latitude 60$^\circ$ in the panel in the first column and third line. This enhancement is also evident in figure~\ref{fig:deflection:hist}. Nitrogen (second column) and iron nuclei (third column) leaving Vir A are largely deviated, arriving at Earth with directions corresponding to large portions of the right hemisphere of the plots for all EGMF models (all lines). It is interesting to note that this region presents an event excess in the data from the Pierre Auger Observatory (figure 4 of reference \citep{Aab_2018}). Nitrogen and iron nuclei leaving For A (green) produce events arriving at Earth with directions concentrated in the latitude range 300$^\circ$--330$^\circ$ for all EGMF models. The anisotropic effect in the arrival directions of UHECR caused by the structure of the EGMF has a large impact in estimate of the dipole direction.

\begin{figure}
  \centering
  \includegraphics[trim=0 250 0 0,clip,width=0.9\columnwidth]{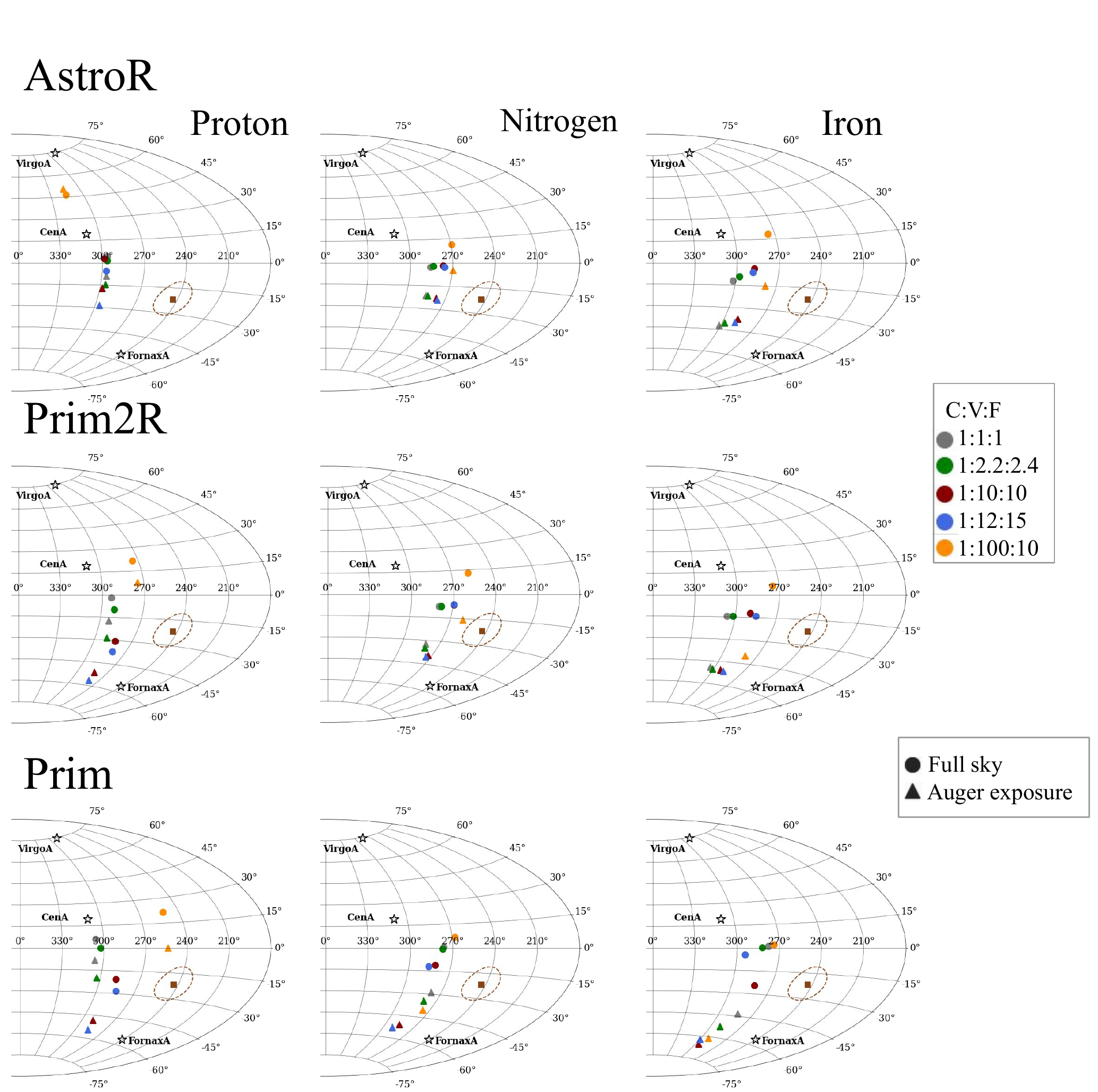}
  \caption{Sky maps in Galactic coordinates and Aitoff projections showing the dipole direction for the simulated events arriving at Earth with energy above 8 EeV. Each line in the figure shows one of the EGMF models considered here. Each column in the figure shows a different nuclei leaving the source: proton, nitrogen, and iron nuclei. Note that in each column of the  figure, all nuclei fragments on the way to Earth are shown as arriving at Earth when only proton, nitrogen, or iron nuclei left the source. The three sources are shown as stars. The brown square shows the direction of the dipole measured by the Pierre Auger Observatory and the dashed brown line shows its one sigma uncertainty. Colored circles show the direction of the dipoles calculated with the simulated events from Cen A, Vir A, and For A. Each color corresponds to a relation of the flux emitted by Cen A : Vir A : For A (C:V:F) as given in the legend.}
  \label{fig:dipole:8:pr_n_fe}
\end{figure}

\begin{figure}
  \centering
  \includegraphics[trim=0 250 0 0,clip,width=0.9\columnwidth]{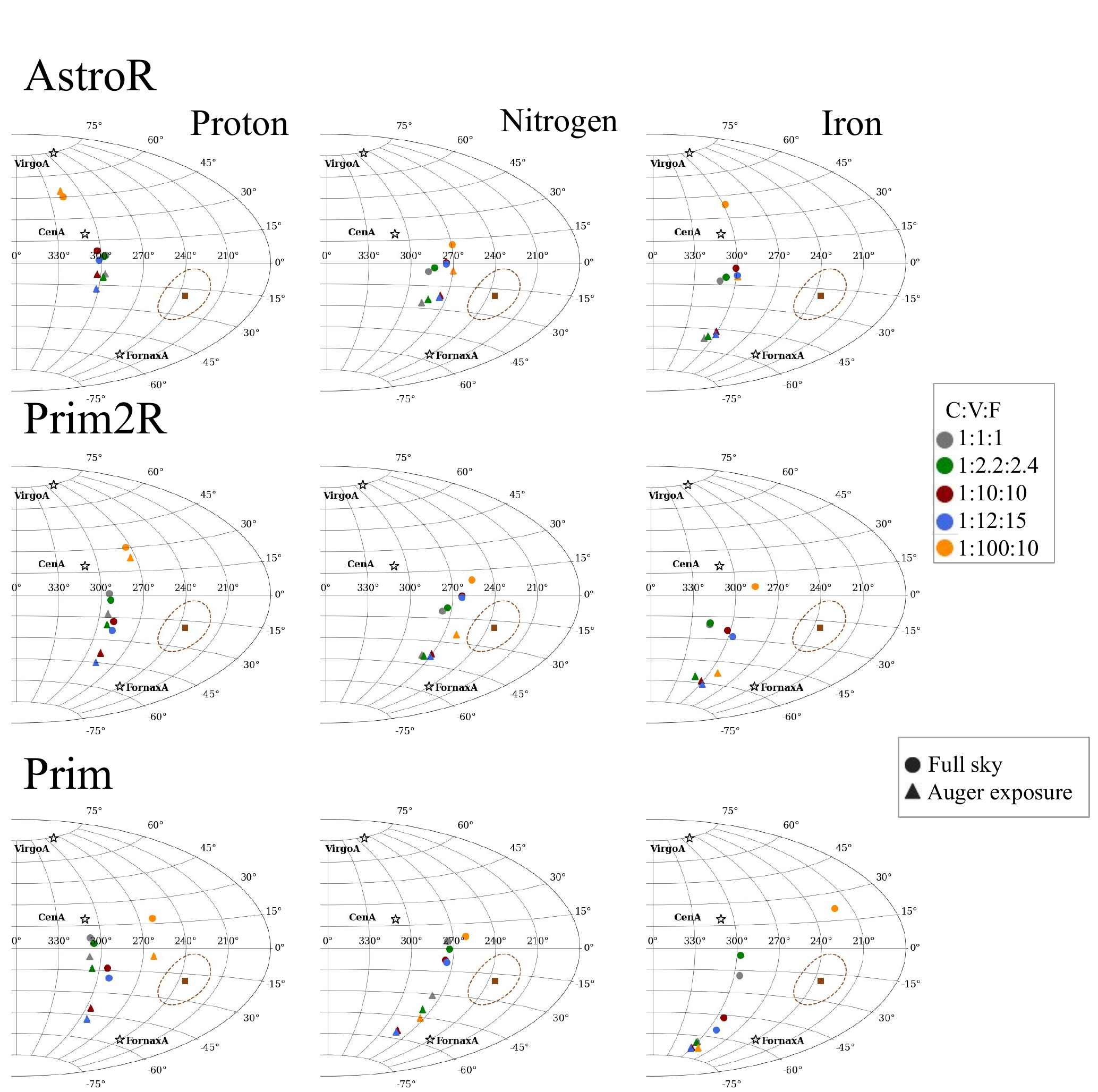}
  \caption{Sky maps in Galactic coordinates and Aitoff projections showing the dipole direction for the simulated events arriving at Earth with energy between 8 and 16 EeV. Each line in the figure shows one of the EGMF models considered here. Each column in the figure shows a different nuclei leaving the source: proton, nitrogen, and iron nuclei. Note that in each column of the  figure, all nuclei fragments on the way to Earth are shown as arriving at Earth when only proton, nitrogen, or iron nuclei left the source. The three sources are shown as stars. The brown square shows the direction of the dipole measured by the Pierre Auger Observatory and the dashed brown line shows its one sigma uncertainty. Colored circles show the direction of the dipoles calculated with the simulated events from Cen A, Vir A, and For A. Each color corresponds to a relation of the flux emitted by Cen A : Vir A : For A (C:V:F) as given in the legend.}
  \label{fig:dipole:8-16:pr_n_fe}
\end{figure}

\begin{figure}
  \centering
  \includegraphics[trim=0 250 0 0,clip,width=0.9\columnwidth]{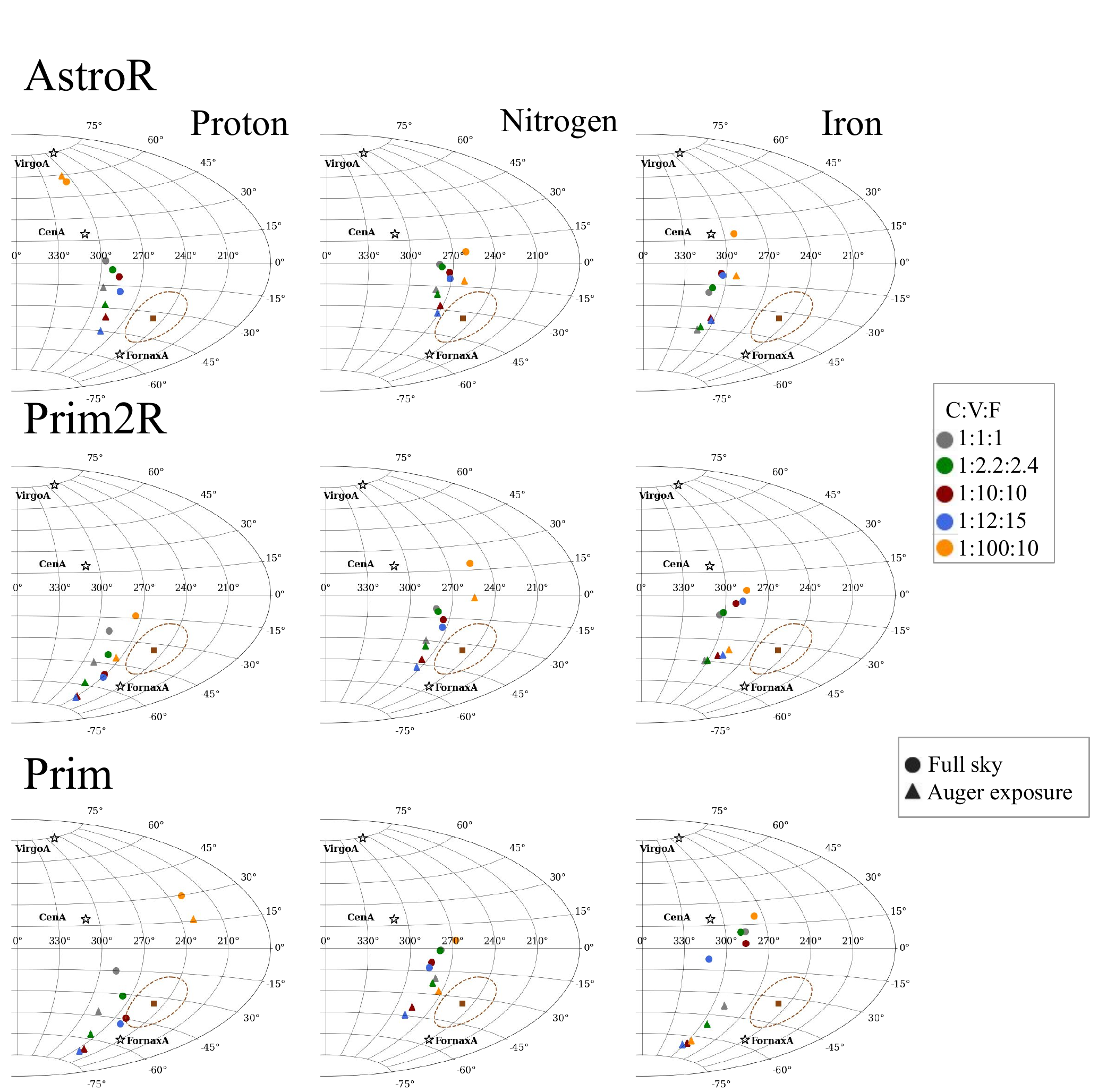}
  \caption{Sky maps in Galactic coordinates and Aitoff projections showing the dipole direction for the simulated events arriving at Earth with energy between 16 and 32 EeV. Each line in the figure shows one of the EGMF models considered here. Each column in the figure shows a different nuclei leaving the source: proton, nitrogen, and iron nuclei. Note that in each column of the  figure, all nuclei fragments on the way to Earth are shown as arriving at Earth when only proton, nitrogen, or iron nuclei left the source. The three sources are shown as stars. The brown square shows the direction of the dipole measured by the Pierre Auger Observatory and the dashed brown line shows its one sigma uncertainty. Colored circles show the direction of the dipoles calculated with the simulated events from Cen A, Vir A, and For A. Each color corresponds to a relation of the flux emitted by Cen A : Vir A : For A (C:V:F) as given in the legend.}
  \label{fig:dipole:16-32:pr_n_fe}
\end{figure}

\begin{figure}
  \centering
  \includegraphics[trim=0 250 0 0,clip,width=0.9\columnwidth]{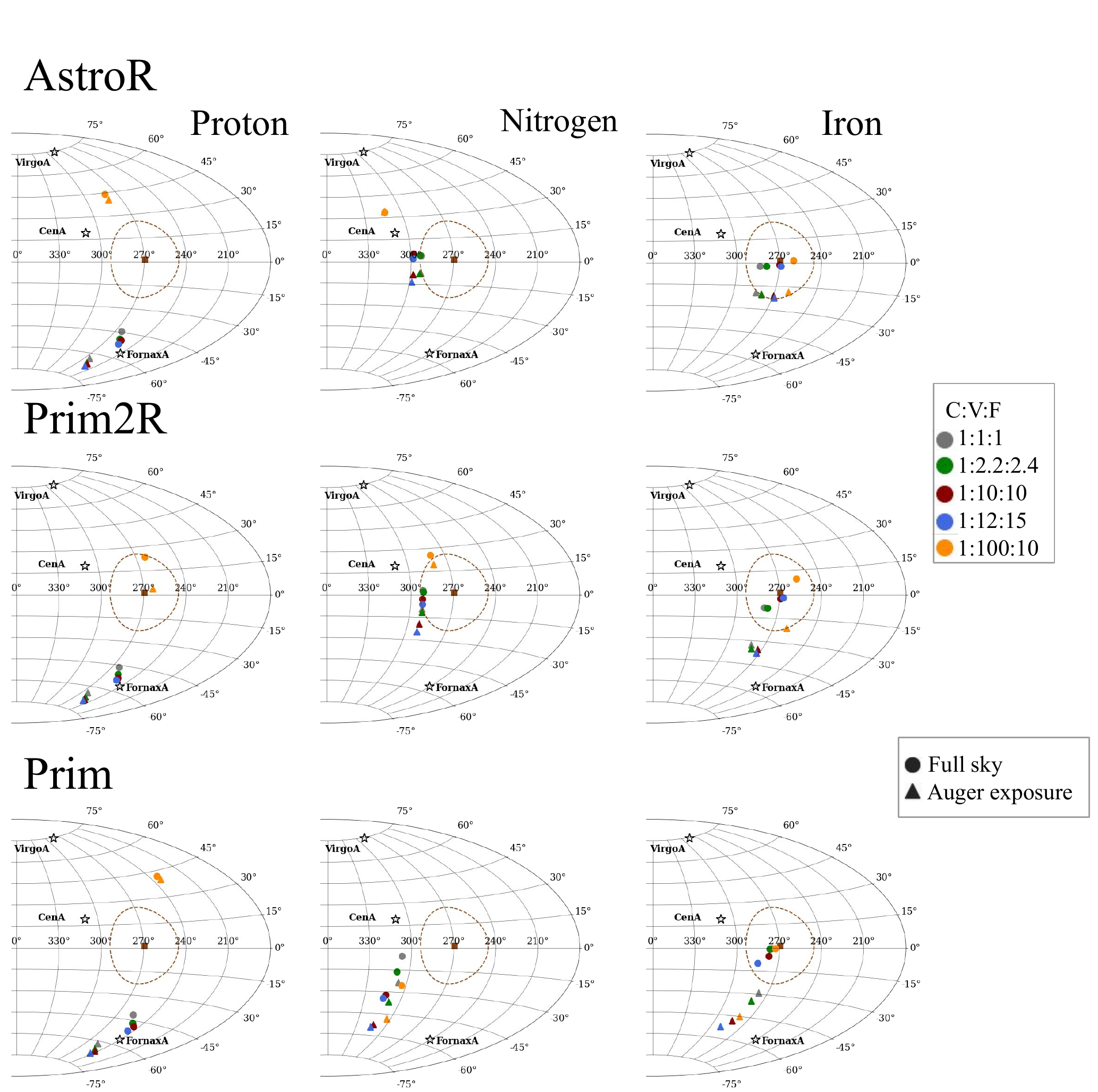}
  \caption{Sky maps in Galactic coordinates and Aitoff projections showing the dipole direction for the simulated events arriving at Earth with energy above 32 EeV. Each line in the figure shows one of the EGMF models considered here. Each column in the figure shows a different nuclei leaving the source: proton, nitrogen, and iron nuclei. Note that in each column of the  figure, all nuclei fragments on the way to Earth are shown as arriving at Earth when only proton, nitrogen, or iron nuclei left the source. The three sources are shown as stars. The brown square shows the direction of the dipole measured by the Pierre Auger Observatory and the dashed brown line shows its one sigma uncertainty. Colored circles show the direction of the dipoles calculated with the simulated events from Cen A, Vir A, and For A. Each color corresponds to a relation of the flux emitted by Cen A : Vir A : For A (C:V:F) as given in the legend.}
  \label{fig:dipole:32:pr_n_fe}
\end{figure}

The dipole direction was calculated using the events arriving at Earth from Cen A, Vir A, and For A with energy $\geq 8$, $8-16$, $16-32$ and $\geq 32$ EeV for several cases (see appendix~\ref{app:dipole}). The calculation was done: a) for each EGMF model, b) for each nucleus type leaving the source, c) for a set of relative output flux for each source, and d) full sky or Auger exposure. Figures~\ref{fig:dipole:8:pr_n_fe}, ~\ref{fig:dipole:8-16:pr_n_fe}, ~\ref{fig:dipole:16-32:pr_n_fe} and~\ref{fig:dipole:32:pr_n_fe} show the calculated directions of the dipole using the simulated events in comparison to the direction of the dipole measured by the Pierre Auger Observatory for the same energy bins. The brown square shows the direction of the dipole measured by the Pierre Auger Observatory, and the dashed brown line shows its one-sigma uncertainty. The one-sigma uncertainty of the simulated events for the same statistics used by the Pierre Auger Observatory is about 2 degrees and therefore was omitted from the plots for the sake of clarity. Proton, nitrogen, and iron nuclei leaving the source are shown in the figures as representations of light, intermediate, and heavy composition leaving the sources. Note that in each column of the figure, all nuclei fragments on the way to Earth are shown as arriving at Earth when only proton, nitrogen, or iron nuclei left the source. Similar plots, leading to the same conclusions for helium and silicon nuclei, are shown in appendix~\ref{app:nuclei}. Colored circles show the direction of the dipoles calculated with the simulated events. Each color corresponds to a relation of the flux emitted by Cen A : Vir A : For A. Five examples of the relation between the fluxes were chosen. Since no other source is considered in this study, only the relative flux is important. Three arbitrary, simplistic and extreme cases of 1:1:1, 1:10:10 and 1:100:10 were calculated. Two scientifically motivated cases are also shown: 1:2.2:2.4 and 1:12:15. The UHECR luminosity of AGNs is considered to be directly related to the mechanical jet power~\citep{10.1046/j.1365-8711.1999.02907.x}. A possible relation between radio luminosity and UHECR flux has been explored in the literature~\citep{Eichmann_2018}. Using the radio luminosity at frequency $\nu = 1.1$ GHz given in section~\ref{sec:method}, the 1:2.2:2.4 ratio is found. Another possible relation between the jet power and X-ray cavity properties present in hot gaseous halos of radio galaxies has been proposed in reference~\citep{10.1093/mnras/stv2712}. Using the X-ray cavity properties of each source considered here~\citep{10.1093/mnras/stv2712}, the 1:12:15 ratio can be calculated. It is not in the scope of this paper to discuss the different estimations of UHECR luminosity for the source. For this reason, a large set of possibilities is used hoping to bracket the true unknown value.

Figures~\ref{fig:dipole:8:pr_n_fe}, ~\ref{fig:dipole:8-16:pr_n_fe}, and ~\ref{fig:dipole:16-32:pr_n_fe} show that the dipole direction measured by the Pierre Auger Observatory in the energy bin $\geq 8$, $8-16$, $16-32$ EeV can not be explained only with the three local sources Cen A, Vir A, and For A by any combination of nuclei leaving the sources independent of the EGMF model be AstroR, Prim2R, or Prim and independent of the ratio between the flux of the sources. Figure~\ref{fig:dipole:32:pr_n_fe} shows that the dipole direction measured by the Pierre Auger Observatory at energies above 32 EeV can be explained only with the three local sources Cen A, Vir A, and For A if they inject heavy nuclei independent of the EGMF model be AstroR, Prim2R, or Prim and independent of the ratio between the fluxes of the sources. The dipole amplitude can not be calculated only with three sources because further sources, even if contributing isotropically to the arrival direction, would decrease the dipole amplitude of nearby sources. For this reason, no comparison between the dipole amplitude measured by the Pierre Auger Observatory and the simulated amplitude with the three sources is shown.

\begin{figure}
  \centering
  \includegraphics[trim=0 200 50 0,clip,width=0.85\columnwidth]{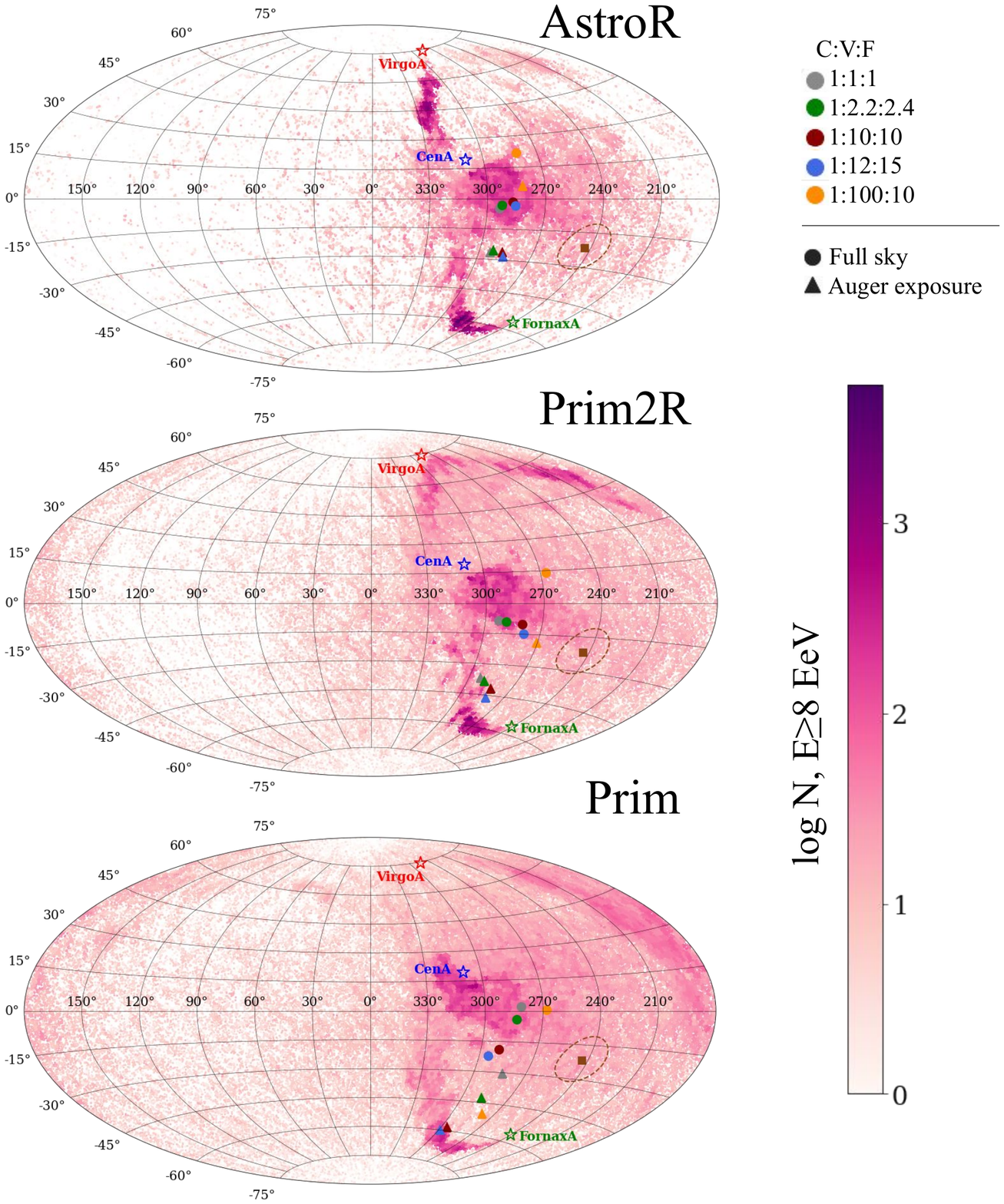}
  \caption{Sky maps in Galactic coordinates and Aitoff projections showing the dipole direction for the simulated events with energy above 8 EeV. Each line in the figure shows one of the EGMF models considered here. The three sources are shown as stars. The brown square shows the direction of the dipole measured by the Pierre Auger Observatory and the dashed brown line shows its one sigma uncertainty. All primaries (H+He+N+Si+Fe) are plotted with equal fluxes leaving the sources. Colored circles show the direction of the dipoles calculated with the simulated events from Cen A, Vir A, and For A. Each color corresponds to a relation of the flux emitted by Cen A : Vir A : For A (C:V:F) as given in the legend.}
  \label{fig:arrival:dipole:8:all}
\end{figure}

\begin{figure}
  \centering
  \includegraphics[trim=0 200 50 0,clip,width=0.85\columnwidth]{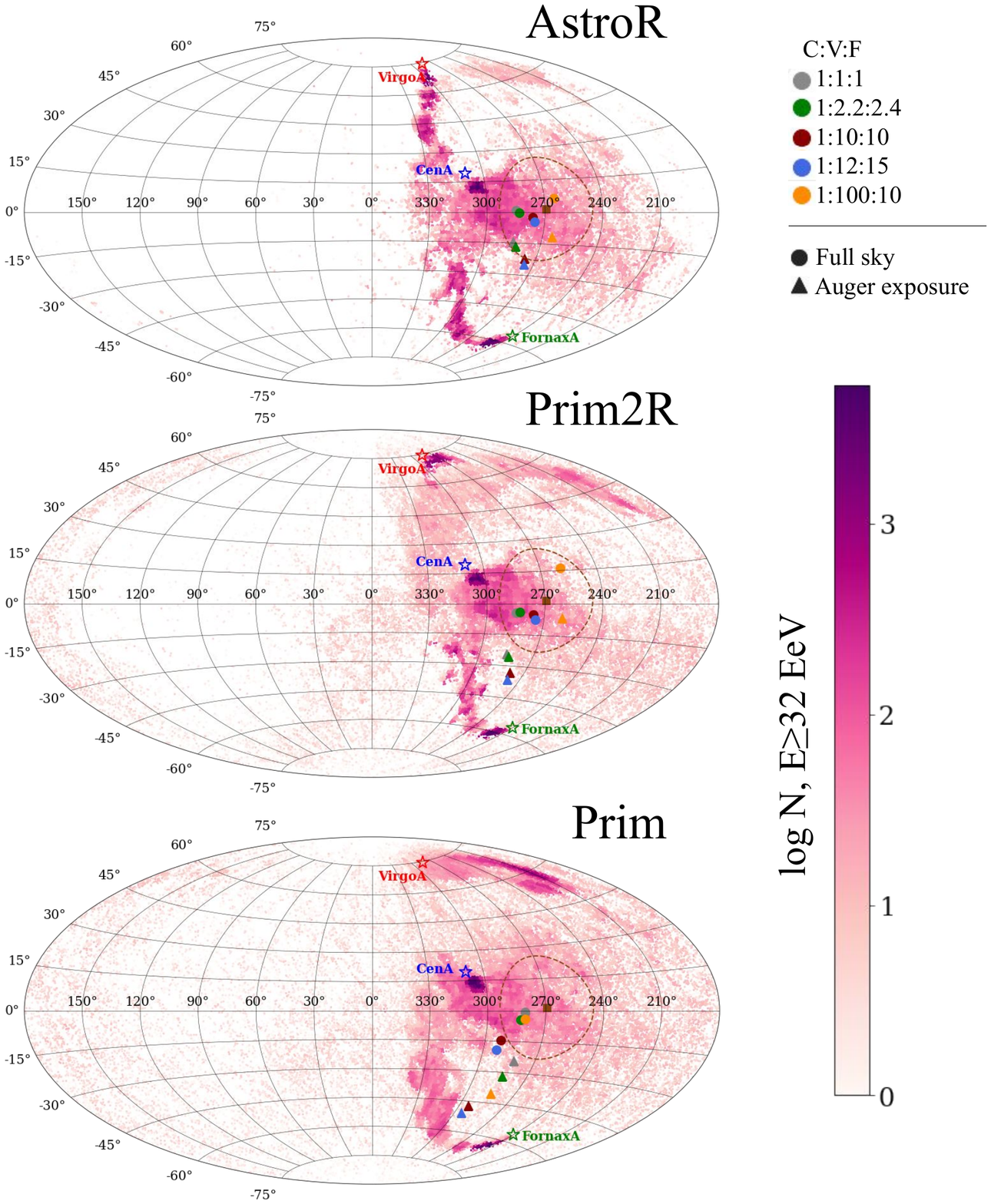}
  \caption{Sky maps in Galactic coordinates and Aitoff projections showing the dipole direction for the simulated events which arrived at Earth with energy above 32 EeV. Each line in the figure shows one of the EGMF models considered here. The three sources are shown as stars. Brown square shows the direction of the dipole measured by the Pierre Auger Observatory and the dashed brown line shows its one sigma uncertainty. All primaries (H+He+N+Si+Fe) are plotted with equal fluxes leaving the sources. Colored circles show the direction of the dipoles. Each color corresponds to a relation of the flux emitted by Cen A : Vir A : For A (C:V:F) as given in the legend.}
  \label{fig:arrival:dipole:32:all}
\end{figure}

Another illustrative view of the contribution of each source is shown in figures~\ref{fig:arrival:dipole:8:all} and ~\ref{fig:arrival:dipole:32:all} in which all nuclei (H+He+N+Si+Fe) are shown with equal fluxes leaving the sources together with the corresponding dipole directions. Other elements of these figures are the same as in figures~\ref{fig:dipole:8:pr_n_fe} and~\ref{fig:dipole:32:pr_n_fe}.

\subsection{The hotspots}
\label{sec:hotspots}

The Pierre Auger Observatory has identified two regions with excess of events in comparison to simulations of isotropic sky for events with energy above 60 EeV~\citep{Aab_2018}. The exact position of the center of these regions, named here hotspots 1 and 2 (HS1 and HS2), were not published by the Pierre Auger Collaboration. From the published maps, the hotspot position was estimated ($l$ = 305$^\circ$, $b$ = 25$^\circ$) (HS1) and ($l$ = 290$^\circ$, $b$ = $-70^\circ$) (HS2). The Telescope Array Observatory identified a cluster of events, named here hotspot 3 (HS3) using 20$^\circ$-radius circles of events with energy above 57 EeV. The hotspot has a Li-Ma statistical significance of 5.1$\sigma$, and is centered around ($l$ = 146$^\circ$.7, $b$ = 43$^\circ$.2) in Galactic coordinates. In this section, the contribution of Cen A, Vir A, and For A to the hotspots under structured EGMF is investigated.

\begin{figure}
  \centering
  \includegraphics[angle=270,width=1.0\columnwidth]{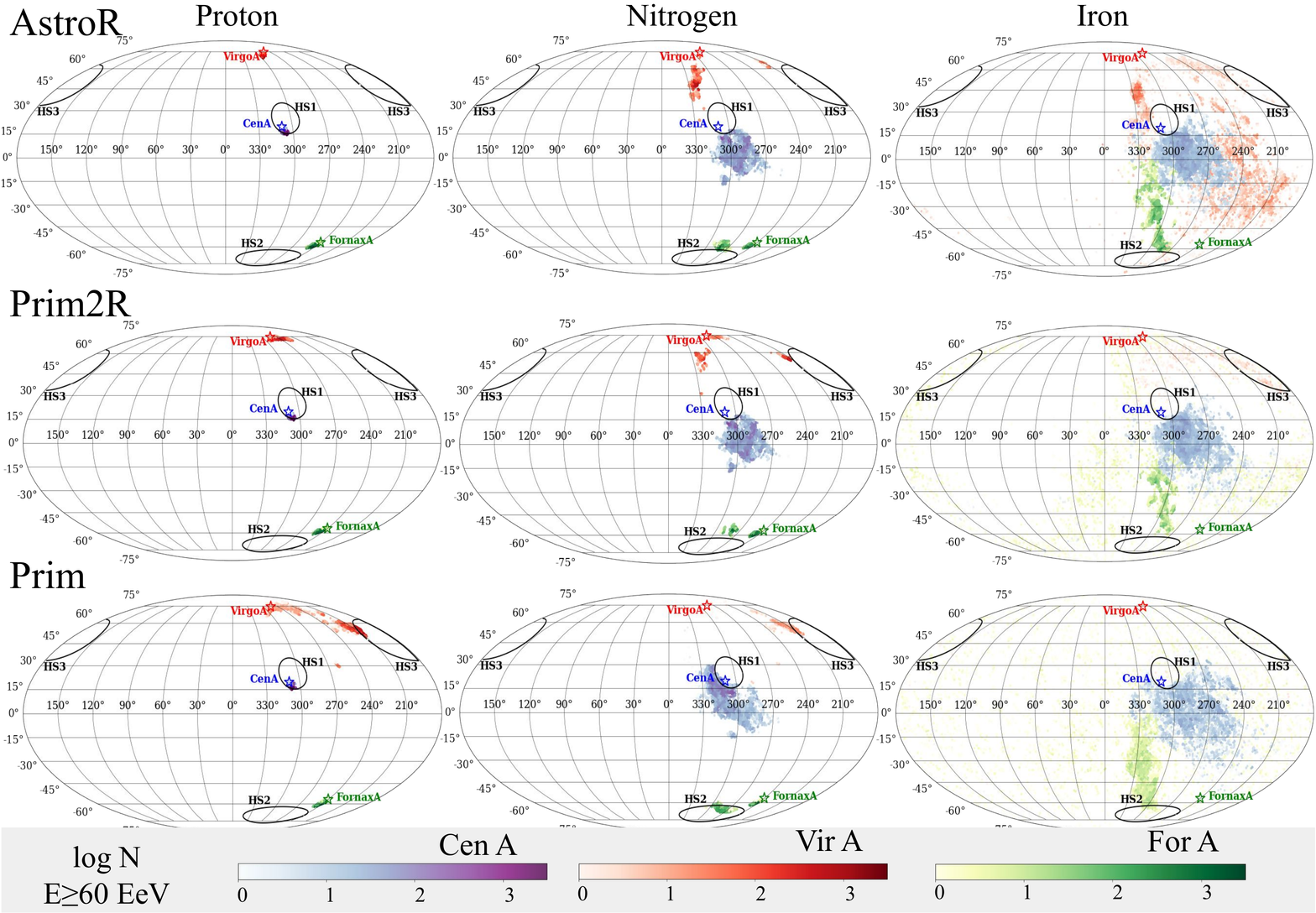}
  \caption{Sky maps in Galactic coordinates and Mollweide projections of all simulated events which arrived at Earth with energy above 60 EeV. Each line in the figure shows one of the EGMF models considered here. Each column in the figure shows a different nuclei leaving the source: proton, nitrogen, and iron nuclei. Note that in each column of the  figure, all nuclei fragments on the way to Earth are shown as arriving at Earth when only proton, nitrogen, or iron nuclei left the source. The three sources are shown as blue, red, and green stars for Cen A, Vir A, and For A, respectively. The flux of events follows the same color-code, each color representing only the events generated in the respective source. The three hotspots regions are circulated by black full lines.}
  \label{fig:arrival:60:pr_n_fe}
\end{figure}

Figure~\ref{fig:arrival:60:pr_n_fe} shows the three hotspot regions circulated by black full lines. It is also shown the arrival directions of the simulated events arriving at Earth with energies above 60 EeV. The HS1 and HS2 were detected for events with energy above 60 EeV, and HS3 was detected for events with energies above 57 EeV. However, the energy reconstruction of both Pierre Auger and Telescope Array Observatories are known to have systematic uncertainties of the order of 14\% and 21\%, respectively~\citep{deligny2020energy}. For this reason, the small difference between the energies above which the hotspots were detected, of the order of 5\%, is neglected in the following studies, and all simulated events with energies above 60 EeV are taken into consideration.

Each line in figure~\ref{fig:arrival:60:pr_n_fe} shows one of the EGMF models considered here. Each column shows a different nucleus leaving the source: proton, nitrogen, and iron. Note that in each column of the figure, all nuclei fragments on the way to Earth are shown as arriving at Earth when only proton, nitrogen, or iron nuclei left the source. Similar plots, leading to the same conclusions for helium and silicon nuclei, are shown in appendix~\ref{app:nuclei}. The three source positions are shown by colored stars: blue, red, and green for Cen A, Vir A, and For A, respectively. The flux of events follows the same color-code, each color representing only the events generated in the respective source. The importance of the structure of the local EGMF is evident in many ways.

Protons leaving all three sources (first column) suffer a negligible deflection in the AstroR and Prim2R EGMF models. However, protons leaving Vir A are deviated in te Prim EGMF model accumulating an excess of events in the HS3 region. Nitrogen nuclei leaving Vir A are also deviated by all EGMF models in the direction of the HS3 region, and specially Prim2R and Prim EGMF models generate a clear accumulation of events in the HS3 region. Iron nuclei leaving Vir A are strongly deviated in all EGMF models. Some accumulation is caused in the AstroR model but isotropic distributions result in Prim2R and Prim. Iron nuclei leaving For A accumulate events along the longitude 330$^\circ$, right above the HS2 region. Nitrogen nuclei accumulate events in the HS2 region for all EGMF models.

\begin{figure}
  \centering
  \includegraphics[trim=0 0 100 0,clip,width=0.65\columnwidth]{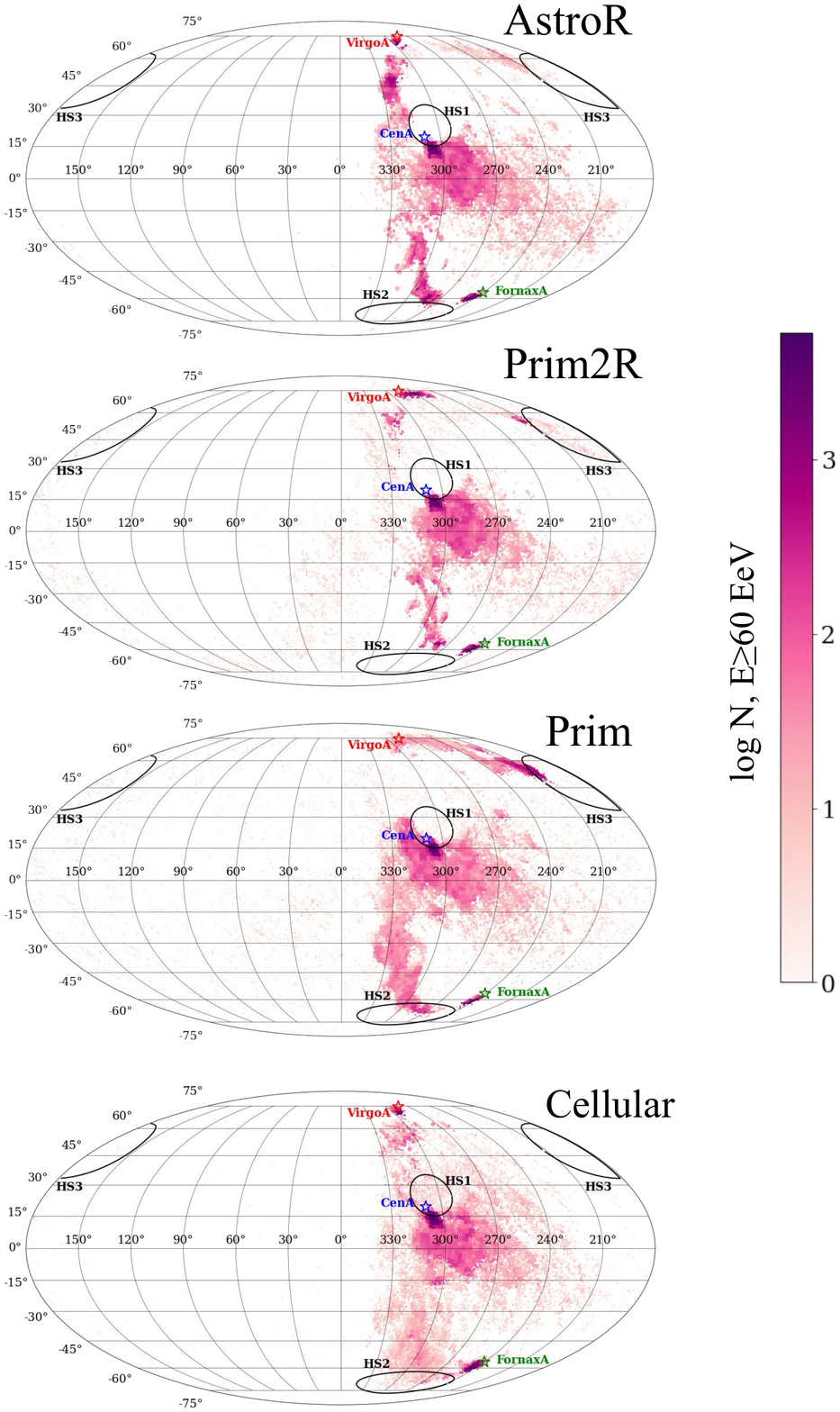}
  \caption{Sky maps in Galactic coordinates and Mollweide projections of all simulated events which arrived at Earth with energy above 60 EeV. Each line in the figure shows one of the EGMF models considered here. The three sources are shown stars. All primaries (H+He+N+Si+Fe) are plotted with equal fluxes leaving the sources. The three hotspots regions are circulated by black full lines.}
  \label{fig:arrival:60:all}
\end{figure}

Another illustrative view of the correlation between HS1 and Cen A, HS2 and For A, and HS3 and Vir A is shown in figure~\ref{fig:arrival:60:all} in which all primaries (H+He+N+Si+Fe) are plotted with equal fluxes leaving the sources. The Cellular EGMF is included for comparison. The structures cause by the AstroR, Prim2R, and Prim EGMF becomes more evident in comparison to the Cellular map. Note that a Cellular EGMF give no contribution to the hotspots. Other elements in this figure are the same as in figure~\ref{fig:arrival:60:pr_n_fe}.

\section{Conclusion}
\label{sec:conclusion}

This paper has two main conclusions as summarized in this section. First, it is shown that the structure of the EGMF dominates the UHECR arrival directions of events from nearby sources. Second, evidences are presented that the measured signals of anisotropy (dipole and hotspots) are largely determined by events generated by Cen A, Vir A, and For A.

The overall effect of the structure of the EGMF on the arrival directions of UHECR originated from nearby sources is very important. At the highest energies investigated here, the galactic magnetic field has a minor contribution to the highly directional effects presented in this conclusion~\citep{ERDMANN201654}. Events are deviated at large angles even for the closest source (see figure~\ref{fig:deflection:hist}). This conclusion confirms previous works~\citep{SIGL2004224,Tanco2001,Medina_Tanco_1998,1995ApJ...455L..21L,PhysRevD.68.043002} with modern tools and EGMF models. It is important to note that these results do not conflict with the finding of reference~\citep{hackstein2018simulations}, in which a continuous source distribution covering a much wider range of distances (up to 100 Mpc) was used. The combination of the two approaches, local point sources and long range continuous source distribution, is a key aspect to determine the influence on the arrival directions of UHECR at Earth. According to the results of this paper and also previous calculations~\citep{Tanco2001,Medina_Tanco_1998}, the immersion of the source in localized EGMF sheets and filaments separated by vast voids has important consequences on the propagation of UHECR. Reference~\citep{Ding_2021} studies imprints of the local scale structure (LSS) of matter in the UHECR sky. Their fundamental hypothesis is that the sources are distributed according to the LSS, and the EGMF is considered to be the simple cellular type. Concerning the hotspots, they concluded that some excess near the South Galactic Pole seen above 38 EeV could potentially be due to the large-scale distribution of matter rather than individual dominant sources. Differently from reference~\citep{Ding_2021}, the calculations shown here suggest that the EGMF structure is the dominant aspect. As summarized below, if a structured EGMF is used in the simulation, several measured signals of anisotropy can be reproduced.

The structure of the EGMF is proposed as the reason for the lack of an excess in the data measured by the Pierre Auger Observatory in the direction of Vir A. In agreement with previous suggestion~\citep{Dolag_2009,Tanco2001}, it was shown in section~\ref{sec:arrival} using the most updated simulation and EGMF models, how the Prim and Prim2R EGMF models creates suppression and enhancement regions in the surroundings of the sources. The AstroR EGMF is much less structured than the others, and the effect is less evident. Earth is in a disfavored position to receive UHECR from Vir A, and the EGMF diverts particles from Vir A away from the Earth direction. Earth is in a favored position to receive UHECR from For A, and the EGMF focuses particles from For A towards Earth.

The suppression/enhancement or lensing effect of the EGMF has important effects on the direction of the dipolar component of the arrival direction distribution. The supression of Vir A, for instance, moves the dipole direction towards lower latitudes. The dipole direction measured by the Pierre Auger Observatory with energy $\geq 8$, $8-16$, $16-32$ EeV can not be explained only by Cen A, Vir A, and For A independently of nuclei leaving the source and of the structure of the EGMF. The dipole direction measured by the Pierre Auger Observatory with energy above 32 EeV can be explained only by Cen A, Vir A, and For A if a large fraction of the emitted UHECR events are heavy nuclei independently of the structure of the EGMF.  This suggests a transition in the predominance from local to distance sources in the dipole signal in the energy range from 8 to 32 EeV.

The EGMF lensing effects is a major inducer of hotspots. The HS1 measured by the Pierre Auger Observatory close to the direction of Cen A is probably caused by light primaries produced in Cen A as has been shown previously and confirmed here. The detailed simulations done here show that the arrival directions of protons under the strongest and most structured EGMF model (Prim) is centered still far away ($\sim10$ degrees) from the centroid of HS1. This implies that HS1 can not be explained only by protons from Cen A. Heavier nuclei from Cen A would not help to populate HS1 because they are deviated away from the HS1 direction. It is also shown in section~\ref{sec:hotspots} that Vir A does not contribute to HS1 even when the strongest and most structured EGMF model is considered. The calculations also show hints that HS2 can receive contribution from intermediate nuclei leaving For A independent of the EGMF model considered. The calculations also show that HS3 can receive contribution from light to intermediate nuclei leaving Vir A if strong and structured EGMF (Prim and Prim2R) are considered. The overall conclusions of this study concerning the hotspots are hints that For A contributes to HS2 and Vir A contributes to HS3, besides the confirmation that Cen A contributes to HS1.

\begin{figure}
  \centering
  \includegraphics[trim=0 200 100 0,clip,width=0.85\columnwidth]{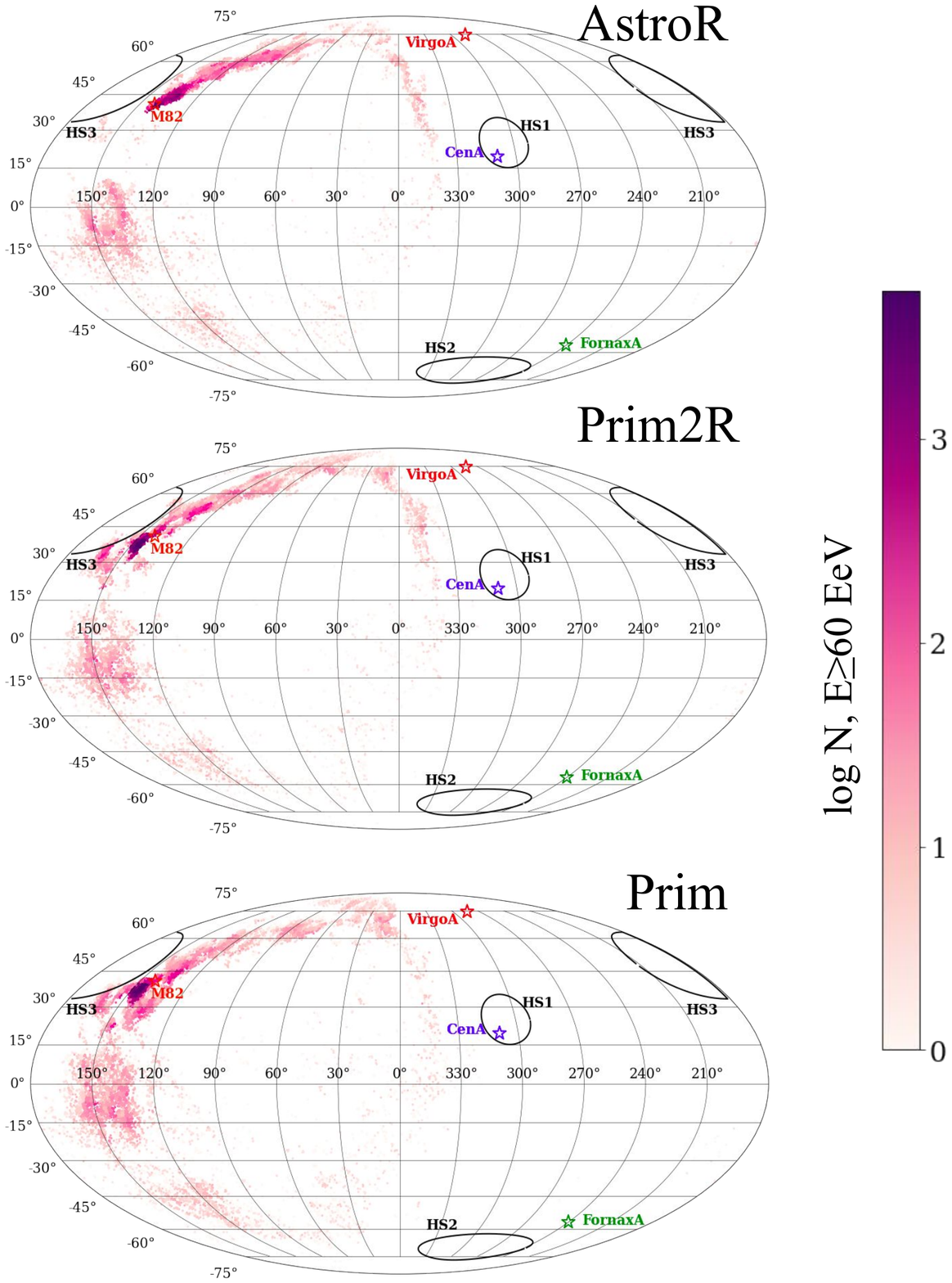}
  \caption{Sky maps in Galactic coordinates and Mollweide projections of all simulated events which arrived at Earth with energy above 60 EeV. Each line in the figure shows one of the EGMF models considered here. The four sources are shown stars. All primaries (H+He+N+Si+Fe) are plotted with equal fluxes leaving the sources. The three hotspots regions are circulated by black full lines. The flux of events arriving at Earth from M82 are shown color-coded according to the legend.}
  \label{fig:maps:arrival:60:m82:all}
\end{figure}

The Telescope Array Collaboration has suggested that HS3 received a strong contribution from the starburst region M82 (NGC 3034)~\citep{Abbasi_2020}. The conclusion of the Telescope Array Collaboration was based on simulation of a cellular type of EGMF that causes a symmetric dispersion of the events around the true source direction. Figure~\ref{fig:maps:arrival:60:m82:all} shows events arriving from M82 when the AstroR, Prim2R, and Prim EGMF models are considered. All nuclei (H+He+N+Si+Fe) are shown with equal fluxes leaving the source. According to these simulations, M82 does not contribute to the HS3 independent of the EGMF model and nuclei type. Under structured EGMF models, the events leaving M82 are diverted asymmetrically and away from HS3.

The EGMF lensing effects shown here represents a paradigm shift from the random walk propagation effect of classical UHECR propagation. The lensing effect proposed previously~\citep{SIGL2004224,Tanco2001,Medina_Tanco_1998,1995ApJ...455L..21L,PhysRevD.68.043002} and studied here with recent EGMF models and modern propagation codes, is shown to be mandatory to interpret the arrival directions measured by the most important observatories. According to the results presented here, the skymap produced by local sources might have large composition-dependent anisotropy that could be tested by observatories with enhanced capabilities to reconstruct the primary particle type of air-showers.

\section*{Acknowledgments}
CO and VdS acknowledge FAPESP Project 2015/15897-1, 2019/10151-2 and 2021/04972-3. The authors acknowledge the National Laboratory for Scientific Computing (LNCC/MCTI,  Brazil) for providing HPC resources of the SDumont supercomputer (http://sdumont.lncc.br). VdS acknowledges CNPq. This study was financed in part by the Coordenação de Aperfeiçoamento de Pessoal de Nível Superior - Brasil (CAPES) - Finance Code 001. Authors thanks Edivaldo Santos Moura and Rog\'erio de Almeida Menezes for useful discussions.

\bibliographystyle{aasjournal}
\bibliography{main.bib}

\begin{thebibliography}{}
\expandafter\ifx\csname natexlab\endcsname\relax\def\natexlab#1{#1}\fi
\providecommand{\url}[1]{\href{#1}{#1}}
\providecommand{\dodoi}[1]{doi:~\href{http://doi.org/#1}{\nolinkurl{#1}}}
\providecommand{\doeprint}[1]{\href{http://ascl.net/#1}{\nolinkurl{http://ascl.net/#1}}}
\providecommand{\doarXiv}[1]{\href{https://arxiv.org/abs/#1}{\nolinkurl{https://arxiv.org/abs/#1}}}

\bibitem[{Aloisio \& Berezinsky(2004)}]{Aloisio_2004}
Aloisio, R., \& Berezinsky, V. 2004, The Astrophysical Journal, 612, 900,
  \dodoi{10.1086/421869}

\bibitem[{{Aublin, J.} \& {Parizot, E.}(2005)}]{parizot_2005}
{Aublin, J.}, \& {Parizot, E.} 2005, Astronomy and Astrophysics, 441, 407,
  \dodoi{10.1051/0004-6361:20052833}

\bibitem[{Batista \& Sigl(2014)}]{Batista_2014}
Batista, R.~A., \& Sigl, G. 2014, Journal of Cosmology and Astroparticle
  Physics, 2014, 031, \dodoi{10.1088/1475-7516/2014/11/031}

\bibitem[{Batista {et~al.}(2016)Batista, Dundovic, Erdmann, Kampert, Kuempel,
  M{\"u}ller, Sigl, van Vliet, Walz, \& Winchen}]{batista2016crpropa}
Batista, R.~A., Dundovic, A., Erdmann, M., {et~al.} 2016, Journal of Cosmology
  and Astroparticle Physics, 2016, 038

\bibitem[{Bell {et~al.}(2018)Bell, Araudo, Matthews, \&
  Blundell}]{bell2018cosmic}
Bell, A., Araudo, A., Matthews, J., \& Blundell, K. 2018, Monthly Notices of
  the Royal Astronomical Society, 473, 2364

\bibitem[{Biermann \& De~Souza(2012)}]{biermann2012centaurus}
Biermann, P.~L., \& De~Souza, V. 2012, The Astrophysical Journal, 746, 72

\bibitem[{Biermann {et~al.}(2010)Biermann, De~Souza, Wiita,
  {et~al.}}]{biermann2010ultra}
Biermann, P.~L., De~Souza, V., Wiita, P.~J., {et~al.} 2010, The Astrophysical
  Journal Letters, 720, L155

\bibitem[{Bretz {et~al.}(2014)Bretz, Erdmann, Schiffer, Walz, \&
  Winchen}]{BRETZ2014110}
Bretz, H.-P., Erdmann, M., Schiffer, P., Walz, D., \& Winchen, T. 2014,
  Astroparticle Physics, 54, 110,
  \dodoi{https://doi.org/10.1016/j.astropartphys.2013.12.002}

\bibitem[{Burns {et~al.}(1983)Burns, Feigelson, \& Schreier}]{burns1983inner}
Burns, J.~O., Feigelson, E., \& Schreier, E. 1983, The Astrophysical Journal,
  273, 128

\bibitem[{{Cavagnolo} {et~al.}(2010){Cavagnolo}, {McNamara}, {Nulsen},
  {Carilli}, {Jones}, \& {B{\^\i}rzan}}]{cavagnolo2010}
{Cavagnolo}, K.~W., {McNamara}, B.~R., {Nulsen}, P.~E.~J., {et~al.} 2010, The
  Astrophysical Journal, 720, 1066, \dodoi{10.1088/0004-637X/720/2/1066}

\bibitem[{Cho \& Ryu(2009)}]{Cho_2009}
Cho, J., \& Ryu, D. 2009, The Astrophysical Journal, 705, L90,
  \dodoi{10.1088/0004-637x/705/1/l90}

\bibitem[{Collaboration {et~al.}(2020)}]{hess2020resolving}
Collaboration, H., {et~al.} 2020, Nature, 582, 356

\bibitem[{Das {et~al.}(2008)Das, Kang, Ryu, \& Cho}]{Das_2008}
Das, S., Kang, H., Ryu, D., \& Cho, J. 2008, The Astrophysical Journal, 682,
  29, \dodoi{10.1086/588278}

\bibitem[{de~Oliveira \& de~Souza(2021)}]{deoliveira2021probing}
de~Oliveira, C., \& de~Souza, V. 2021, The European Physical Journal C, 81, 1

\bibitem[{Dermer \& Razzaque(2010)}]{Dermer_2010}
Dermer, C.~D., \& Razzaque, S. 2010, The Astrophysical Journal, 724, 1366,
  \dodoi{10.1088/0004-637x/724/2/1366}

\bibitem[{Ding {et~al.}(2021)Ding, Globus, \& Farrar}]{Ding_2021}
Ding, C., Globus, N., \& Farrar, G.~R. 2021, The Astrophysical Journal Letters,
  913, L13, \dodoi{10.3847/2041-8213/abf11e}

\bibitem[{Dolag {et~al.}(2009)Dolag, Kachelrie{\ss}, \& Semikoz}]{Dolag_2009}
Dolag, K., Kachelrie{\ss}, M., \& Semikoz, D. 2009, Journal of Cosmology and
  Astroparticle Physics, 2009, 033, \dodoi{10.1088/1475-7516/2009/01/033}

\bibitem[{Durrer \& Neronov(2013)}]{durrer}
Durrer, R., \& Neronov, A. 2013, Astron Astrophys Rev, 21,
  \dodoi{/10.1007/s00159-013-0062-7}

\bibitem[{Eckert {et~al.}(2015)}]{density}
Eckert, D., {et~al.} 2015, Nature, 528, 105

\bibitem[{Eichmann(2019)}]{eichmann2019high}
Eichmann, B. 2019, Journal of Cosmology and Astroparticle Physics, 2019, 009

\bibitem[{Eichmann {et~al.}(2018)Eichmann, Rachen, Merten, van Vliet, \&
  Tjus}]{Eichmann_2018}
Eichmann, B., Rachen, J., Merten, L., van Vliet, A., \& Tjus, J.~B. 2018,
  Journal of Cosmology and Astroparticle Physics, 2018, 036,
  \dodoi{10.1088/1475-7516/2018/02/036}

\bibitem[{Erdmann {et~al.}(2016)Erdmann, Müller, Urban, \&
  Wirtz}]{ERDMANN201654}
Erdmann, M., Müller, G., Urban, M., \& Wirtz, M. 2016, Astroparticle Physics,
  85, 54, \dodoi{https://doi.org/10.1016/j.astropartphys.2016.10.002}

\bibitem[{Farrar {et~al.}(2013)Farrar, Jansson, Feain, \&
  Gaensler}]{Farrar_2013}
Farrar, G.~R., Jansson, R., Feain, I.~J., \& Gaensler, B. 2013, Journal of
  Cosmology and Astroparticle Physics, 2013, 023,
  \dodoi{10.1088/1475-7516/2013/01/023}

\bibitem[{Feigelson {et~al.}(1981)Feigelson, Schreier, Delvaille, Giacconi,
  Grindlay, \& Lightman}]{feigelson1981x}
Feigelson, E., Schreier, E., Delvaille, J., {et~al.} 1981, The Astrophysical
  Journal, 251, 31

\bibitem[{Ferrarese {et~al.}(2006)Ferrarese, Cote, Jordan, Peng, Blakeslee,
  Piatek, Mei, Merritt, Milosavljevi{\'{c}}, Tonry, \& West}]{Ferrarese_2006}
Ferrarese, L., Cote, P., Jordan, A., {et~al.} 2006, The Astrophysical Journal
  Supplement Series, 164, 334, \dodoi{10.1086/501350}

\bibitem[{Ghisellini {et~al.}(2008)Ghisellini, Ghirlanda, Tavecchio,
  Fraternali, \& Pareschi}]{10.1111/j.1745-3933.2008.00547.x}
Ghisellini, G., Ghirlanda, G., Tavecchio, F., Fraternali, F., \& Pareschi, G.
  2008, Monthly Notices of the Royal Astronomical Society: Letters, 390, L88,
  \dodoi{10.1111/j.1745-3933.2008.00547.x}

\bibitem[{Gilmore {et~al.}(2012)Gilmore, Somerville, Primack, \&
  Dom{\'\i}nguez}]{gilmore2012semi}
Gilmore, R.~C., Somerville, R.~S., Primack, J.~R., \& Dom{\'\i}nguez, A. 2012,
  Monthly Notices of the Royal Astronomical Society, 422, 3189

\bibitem[{Ginzburg \& Syrovatskii(1963)}]{ginzburg1963cosmic}
Ginzburg, V., \& Syrovatskii, S. 1963, Soviet Astronomy, 7, 357

\bibitem[{Godfrey \& Shabala(2016)}]{godfrey2016mutual}
Godfrey, L., \& Shabala, S. 2016, Monthly Notices of the Royal Astronomical
  Society, 456, 1172

\bibitem[{Godfrey \& Shabala(2015)}]{10.1093/mnras/stv2712}
Godfrey, L. E.~H., \& Shabala, S.~S. 2015, Monthly Notices of the Royal
  Astronomical Society, 456, 1172, \dodoi{10.1093/mnras/stv2712}

\bibitem[{Greisen(1966)}]{Greisen1966}
Greisen, K. 1966, Phys. Rev. Lett., 16, 748.
\newblock \url{http://link.aps.org/abstract/PRL/v16/p748}

\bibitem[{Hackstein {et~al.}(2018)Hackstein, Vazza, Br{\"u}ggen, Sorce, \&
  Gottl{\"o}ber}]{hackstein2018simulations}
Hackstein, S., Vazza, F., Br{\"u}ggen, M., Sorce, J.~G., \& Gottl{\"o}ber, S.
  2018, Monthly Notices of the Royal Astronomical Society, 475, 2519

\bibitem[{Hackstein {et~al.}(2016)Hackstein, Vazza, Brüggen, Sigl, \&
  Dundovic}]{10.1093/mnras/stw1903}
Hackstein, S., Vazza, F., Brüggen, M., Sigl, G., \& Dundovic, A. 2016, Monthly
  Notices of the Royal Astronomical Society, 462, 3660,
  \dodoi{10.1093/mnras/stw1903}

\bibitem[{{Hada, Kazuhiro}(2013)}]{Hada_2013}
{Hada, Kazuhiro}. 2013, EPJ Web of Conferences, 61, 01002,
  \dodoi{10.1051/epjconf/20136101002}

\bibitem[{Halzen \& Hooper(2002)}]{Halzen_2002}
Halzen, F., \& Hooper, D. 2002, Reports on Progress in Physics, 65, 1025,
  \dodoi{10.1088/0034-4885/65/7/201}

\bibitem[{Hardcastle {et~al.}(2003)Hardcastle, Worrall, Kraft, Forman, Jones,
  \& Murray}]{hardcastle2003radio}
Hardcastle, M., Worrall, D., Kraft, R., {et~al.} 2003, The Astrophysical
  Journal, 593, 169

\bibitem[{Hillas(1984)}]{doi:10.1146/annurev.aa.22.090184.002233}
Hillas, A.~M. 1984, Annual Review of Astronomy and Astrophysics, 22, 425,
  \dodoi{10.1146/annurev.aa.22.090184.002233}

\bibitem[{Israel(1998)}]{cena00}
Israel, F. 1998, The Astronomy and Astrophysics Review, 8, 237.
\newblock \url{https://doi.org/10.1007/s001590050011}

\bibitem[{Jansson \& Farrar(2012)}]{Jansson_2012}
Jansson, R., \& Farrar, G.~R. 2012, The Astrophysical Journal, 757, 14,
  \dodoi{10.1088/0004-637x/757/1/14}

\bibitem[{Jensen {et~al.}(2003)Jensen, Tonry, Barris, Thompson, Liu, Rieke,
  Ajhar, \& Blakeslee}]{Jensen_2003}
Jensen, J.~B., Tonry, J.~L., Barris, B.~J., {et~al.} 2003, The Astrophysical
  Journal, 583, 712, \dodoi{10.1086/345430}

\bibitem[{Kachelriess {et~al.}(2009)Kachelriess, Ostapchenko, \&
  Tomas}]{kachelriess2009high}
Kachelriess, M., Ostapchenko, S., \& Tomas, R. 2009, New Journal of Physics,
  11, 065017

\bibitem[{{Kino, M.} {et~al.}(2013){Kino, M.}, {Takahara, F.}, {Hada, K.}, \&
  {Doi, A.}}]{Kino_2013}
{Kino, M.}, {Takahara, F.}, {Hada, K.}, \& {Doi, A.} 2013, EPJ Web of
  Conferences, 61, 01009, \dodoi{10.1051/epjconf/20136101009}

\bibitem[{Kobzar {et~al.}(2019)Kobzar, Hnatyk, Marchenko, \&
  Sushchov}]{10.1093/mnras/stz094}
Kobzar, O., Hnatyk, B., Marchenko, V., \& Sushchov, O. 2019, Monthly Notices of
  the Royal Astronomical Society, 484, 1790, \dodoi{10.1093/mnras/stz094}

\bibitem[{Kotera \& Olinto(2011)}]{kotera2011astrophysics}
Kotera, K., \& Olinto, A.~V. 2011, Annual Review of Astronomy and Astrophysics,
  49, 119

\bibitem[{Kraft {et~al.}(2002)Kraft, Forman, Jones, Murray, Hardcastle, \&
  Worrall}]{Kraft_2002}
Kraft, R.~P., Forman, W.~R., Jones, C., {et~al.} 2002, The Astrophysical
  Journal, 569, 54, \dodoi{10.1086/339062}

\bibitem[{Kronberg(1994)}]{Kronberg_1994}
Kronberg, P.~P. 1994, Reports on Progress in Physics, 57, 325,
  \dodoi{10.1088/0034-4885/57/4/001}

\bibitem[{Lang {et~al.}(2020)Lang, Taylor, Ahlers, \&
  de~Souza}]{lang2020revisiting}
Lang, R.~G., Taylor, A.~M., Ahlers, M., \& de~Souza, V. 2020, Physical Review
  D, 102, 063012

\bibitem[{Lang {et~al.}(2021)Lang, Taylor, \& de~Souza}]{PhysRevD.103.063005}
Lang, R.~G., Taylor, A.~M., \& de~Souza, V. 2021, Phys. Rev. D, 103, 063005,
  \dodoi{10.1103/PhysRevD.103.063005}

\bibitem[{{Lee} {et~al.}(1995){Lee}, {Olinto}, \& {Sigl}}]{1995ApJ...455L..21L}
{Lee}, S., {Olinto}, A.~V., \& {Sigl}, G. 1995, The Astrophysical Journal
  Letter, 455, L21, \dodoi{10.1086/309812}

\bibitem[{{Maccagni, F. M.} {et~al.}(2020){Maccagni, F. M.}, {Murgia, M.},
  {Serra, P.}, {Govoni, F.}, {Morokuma-Matsui, K.}, {Kleiner, D.}, {Buchner,
  S.}, {J\'ozsa, G. I. G.}, {Kamphuis, P.}, {Makhathini, S.}, {Moln\'ar, D.
  Cs.}, {Prokhorov, D. A.}, {Ramaila, A.}, {Ramatsoku, M.}, {Thorat, K.}, \&
  {Smirnov, O.}}]{Maccagni_2020}
{Maccagni, F. M.}, {Murgia, M.}, {Serra, P.}, {et~al.} 2020, Astronomy and
  Astrophysics, 634, A9, \dodoi{10.1051/0004-6361/201936867}

\bibitem[{Macchetto {et~al.}(1997)Macchetto, Marconi, Axon, Capetti, Sparks, \&
  Crane}]{Macchetto_1997}
Macchetto, F., Marconi, A., Axon, D.~J., {et~al.} 1997, The Astrophysical
  Journal, 489, 579, \dodoi{10.1086/304823}

\bibitem[{{Mannheim} \& {Biermann}(1989)}]{1989A&A...221..211M}
{Mannheim}, K., \& {Biermann}, P.~L. 1989, Astronomy\&Astrophysics, 221, 211

\bibitem[{Matthews {et~al.}(2018)Matthews, Bell, Blundell, \&
  Araudo}]{matthews2018fornax}
Matthews, J.~H., Bell, A.~R., Blundell, K.~M., \& Araudo, A.~T. 2018, Monthly
  Notices of the Royal Astronomical Society: Letters, 479, L76

\bibitem[{Matthews {et~al.}(2019)Matthews, Bell, Blundell, \&
  Araudo}]{matthews2019ultrahigh}
---. 2019, Monthly Notices of the Royal Astronomical Society, 482, 4303

\bibitem[{{Norman} {et~al.}(1995){Norman}, {Melrose}, \&
  {Achterberg}}]{1995ApJ...454...60N}
{Norman}, C.~A., {Melrose}, D.~B., \& {Achterberg}, A. 1995, The Astrophysical
  Journal, 454, 60, \dodoi{10.1086/176465}

\bibitem[{Nowak {et~al.}(2008)Nowak, Saglia, Thomas, Bender, Davies, \&
  Gebhardt}]{10.1111/j.1365-2966.2008.13960.x}
Nowak, N., Saglia, R.~P., Thomas, J., {et~al.} 2008, Monthly Notices of the
  Royal Astronomical Society, 391, 1629,
  \dodoi{10.1111/j.1365-2966.2008.13960.x}

\bibitem[{{O. Deligny for the Pierre Auger and Telescope Array
  Collaborations}(2019)}]{deligny2020energy}
{O. Deligny for the Pierre Auger and Telescope Array Collaborations}. 2019,
  Proceedings of 36th International Cosmic Ray Conference, 234.
\newblock \doarXiv{2001.08811}

\bibitem[{Olinto(2000)}]{OLINTO2000329}
Olinto, A. 2000, Physics Reports, 333-334, 329,
  \dodoi{https://doi.org/10.1016/S0370-1573(00)00028-4}

\bibitem[{Pacholczyk(1970)}]{pacholczyk1970radio}
Pacholczyk, A.~G. 1970, Radio astrophysics: {N}onthermal processes in galactic
  and extragalactic sources (San Francisco, U.S.A.: W.H. Freeman \& Co Ltd.)

\bibitem[{Rachen(2008)}]{rachen2008ultrahigh}
Rachen, J.~P. 2008, Ultra-high energy cosmic rays from radio galaxies
  revisited.
\newblock \doarXiv{0808.0349}

\bibitem[{Rachen \& Biermann(1993)}]{rachen1993extragalactic}
Rachen, J.~P., \& Biermann, P.~L. 1993, Astronomy and Astrophysics, 272, 161

\bibitem[{Rachen \& M\'esz\'aros(1998)}]{PhysRevD.58.123005}
Rachen, J.~P., \& M\'esz\'aros, P. 1998, Phys. Rev. D, 58, 123005,
  \dodoi{10.1103/PhysRevD.58.123005}

\bibitem[{{Rejkuba, M.}(2004)}]{refId0}
{Rejkuba, M.} 2004, Astronomy and Astrophysics, 413, 903,
  \dodoi{10.1051/0004-6361:20034031}

\bibitem[{Romero {et~al.}(1996)Romero, Combi, {Perez Bergliaffa}, \&
  Anchordoqui}]{ROMERO1996279}
Romero, G.~E., Combi, J.~A., {Perez Bergliaffa}, S.~E., \& Anchordoqui, L.~A.
  1996, Astroparticle Physics, 5, 279,
  \dodoi{https://doi.org/10.1016/0927-6505(96)00029-1}

\bibitem[{R\"oser \& Meisenheimer(1997)}]{m87book}
R\"oser, H.-J., \& Meisenheimer, K., eds. 1997, {The Radio Galaxy Messier 87}
  (Springer), \dodoi{10.1007/BFb0106412}

\bibitem[{Ryu {et~al.}(2008)Ryu, Kang, Cho, \& Das}]{Ryu909}
Ryu, D., Kang, H., Cho, J., \& Das, S. 2008, Science, 320, 909,
  \dodoi{10.1126/science.1154923}

\bibitem[{{Schweizer}(1980)}]{1980ApJ...237..303S}
{Schweizer}, F. 1980, The Astrophysical Journal, 237, 303,
  \dodoi{10.1086/157870}

\bibitem[{Sigl {et~al.}(2003)Sigl, Miniati, \& Ensslin}]{PhysRevD.68.043002}
Sigl, G., Miniati, F., \& Ensslin, T.~A. 2003, Phys. Rev. D, 68, 043002,
  \dodoi{10.1103/PhysRevD.68.043002}

\bibitem[{Sigl {et~al.}(2004)Sigl, Miniati, \& Ensslin}]{SIGL2004224}
---. 2004, Nuclear Physics B - Proceedings Supplements, 136, 224,
  \dodoi{https://doi.org/10.1016/j.nuclphysbps.2004.10.043}

\bibitem[{Snios {et~al.}(2019)Snios, Nulsen, Kraft, Cheung, Meyer, Forman,
  Jones, \& Murray}]{Snios_2019}
Snios, B., Nulsen, P. E.~J., Kraft, R.~P., {et~al.} 2019, The Astrophysical
  Journal, 879, 8, \dodoi{10.3847/1538-4357/ab2119}

\bibitem[{Subramanian(2016)}]{Subramanian_2016}
Subramanian, K. 2016, Reports on Progress in Physics, 79, 076901,
  \dodoi{10.1088/0034-4885/79/7/076901}

\bibitem[{{Sun, Xiao-na} {et~al.}(2016){Sun, Xiao-na}, {Yang, Rui-zhi},
  {Mckinley, Benjamin}, \& {Aharonian, Felix}}]{Sun_2016}
{Sun, Xiao-na}, {Yang, Rui-zhi}, {Mckinley, Benjamin}, \& {Aharonian, Felix}.
  2016, Astronomy and Astrophysics, 595, A29,
  \dodoi{10.1051/0004-6361/201629069}

\bibitem[{Supanitsky \& de~Souza(2013)}]{supanitsky2013upper}
Supanitsky, A.~D., \& de~Souza, V. 2013, Journal of Cosmology and Astroparticle
  Physics, 2013, 023

\bibitem[{Tanco(1998)}]{Medina_Tanco_1998}
Tanco, G. A.~M. 1998, The Astrophysical Journal, 505, L79,
  \dodoi{10.1086/311615}

\bibitem[{Tanco(2001)}]{Tanco2001}
Tanco, G.~M. 2001, Cosmic Magnetic Fields from the Perspective of
  Ultra-High-Energy Cosmic Rays Propagation, ed. M.~Lemoine \& G.~Sigl (Berlin,
  Heidelberg: Springer Berlin Heidelberg), 155--180,
  \dodoi{10.1007/3-540-45615-5_7}

\bibitem[{Taylor {et~al.}(2011)Taylor, Ahlers, \& Aharonian}]{taylor2011need}
Taylor, A.~M., Ahlers, M., \& Aharonian, F.~A. 2011, Physical Review D, 84,
  105007

\bibitem[{{The Pierre Auger Collaboration}(2015)}]{2015172}
{The Pierre Auger Collaboration}. 2015, Nuclear Instruments and Methods in
  Physics Research Section A: Accelerators, Spectrometers, Detectors and
  Associated Equipment, 798, 172,
  \dodoi{https://doi.org/10.1016/j.nima.2015.06.058}

\bibitem[{{The Pierre Auger Collaboration}(2017)}]{1266}
---. 2017, Science, 357, 1266, \dodoi{10.1126/science.aan4338}

\bibitem[{{The Pierre Auger
  Collaboration}(2018{\natexlab{a}})}]{aab2018indication}
---. 2018{\natexlab{a}}, The Astrophysical Journal Letters, 853, L29

\bibitem[{{The Pierre Auger Collaboration}(2018{\natexlab{b}})}]{Aab_2018}
---. 2018{\natexlab{b}}, The Astrophysical Journal, 868, 4,
  \dodoi{10.3847/1538-4357/aae689}

\bibitem[{{The Pierre Auger Collaboration}(2021)}]{deAlmeida:20212Z}
{The Pierre Auger Collaboration}. 2021, in Proceedings of 37th International
  Cosmic Ray Conference {\textemdash} PoS(ICRC2021), Vol. 395, 335,
  \dodoi{10.22323/1.395.0335}

\bibitem[{{The Pierre Auger Observatory}(2020)}]{Aab_2020}
{The Pierre Auger Observatory}. 2020, Journal of Instrumentation, 15, P10021,
  \dodoi{10.1088/1748-0221/15/10/p10021}

\bibitem[{{The Telescope Array Collaboration}(2008)}]{KAWAI2008221}
{The Telescope Array Collaboration}. 2008, Nuclear Physics B - Proceedings
  Supplements, 175-176, 221,
  \dodoi{https://doi.org/10.1016/j.nuclphysbps.2007.11.002}

\bibitem[{{The Telescope Array Collaboration}(2014)}]{Abbasi_2014}
---. 2014, The Astrophysical Journal, 790, L21,
  \dodoi{10.1088/2041-8205/790/2/l21}

\bibitem[{{The Telescope Array Collaboration}(2020)}]{Abbasi_2020}
---. 2020, The Astrophysical Journal, 899, 86, \dodoi{10.3847/1538-4357/aba26c}

\bibitem[{van Velzen {et~al.}(2012)van Velzen, Falcke, Schellart,
  Nierstenh{\"o}fer, \& Kampert}]{van2012radio}
van Velzen, S., Falcke, H., Schellart, P., Nierstenh{\"o}fer, N., \& Kampert,
  K.-H. 2012, Astronomy \& Astrophysics, 544, A18

\bibitem[{{Vietri}(1995)}]{1995ApJ...453..883V}
{Vietri}, M. 1995, The Astrophysical Journal, 453, 883, \dodoi{10.1086/176448}

\bibitem[{Wang {et~al.}(2008)Wang, Razzaque, \&
  M{\'{e}}sz{\'{a}}ros}]{Wang_2008}
Wang, X.-Y., Razzaque, S., \& M{\'{e}}sz{\'{a}}ros, P. 2008, The Astrophysical
  Journal, 677, 432, \dodoi{10.1086/529018}

\bibitem[{Waxman(1995)}]{PhysRevLett.75.386}
Waxman, E. 1995, Phys. Rev. Lett., 75, 386, \dodoi{10.1103/PhysRevLett.75.386}

\bibitem[{Waxman \& Miralda-Escudé(1996)}]{Waxman_1996}
Waxman, E., \& Miralda-Escudé, J. 1996, The Astrophysical Journal, 472,
  L89–L92, \dodoi{10.1086/310367}

\bibitem[{Widrow {et~al.}(2012)Widrow, Ryu, Schleicher, Subramanian, Tsagas, \&
  Treumann}]{widrow_2012}
Widrow, L.~M., Ryu, D., Schleicher, D.~R., {et~al.} 2012, Space Science
  Reviews, 166, 37

\bibitem[{Willott {et~al.}(1999)Willott, Rawlings, Blundell, \&
  Lacy}]{10.1046/j.1365-8711.1999.02907.x}
Willott, C.~J., Rawlings, S., Blundell, K.~M., \& Lacy, M. 1999, Monthly
  Notices of the Royal Astronomical Society, 309, 1017,
  \dodoi{10.1046/j.1365-8711.1999.02907.x}

\bibitem[{Wykes {et~al.}(2013)Wykes, Croston, Hardcastle, Eilek, Biermann,
  Achterberg, Bray, Lazarian, Haverkorn, Protheroe, {et~al.}}]{wykes2013mass}
Wykes, S., Croston, J.~H., Hardcastle, M.~J., {et~al.} 2013, Astronomy \&
  Astrophysics, 558, A19

\bibitem[{Yang {et~al.}(2012)Yang, Sahakyan, de~Ona~Wilhelmi, Aharonian, \&
  Rieger}]{yang2012deep}
Yang, R.-Z., Sahakyan, N., de~Ona~Wilhelmi, E., Aharonian, F., \& Rieger, F.
  2012, Astronomy \& Astrophysics, 542, A19

\bibitem[{Zatsepin \& Kuz'min(1966)}]{bib:zk}
Zatsepin, G.~T., \& Kuz'min, V.~A. 1966, Soviet Journal of Experimental and
  Theoretical Physics Letters, 4, 78

\end{thebibliography}

\appendix

\section{UHECR emission model}
\label{app:emission}

In this appendix, the UHECR emission model used in this paper is explained. First-order Fermi acceleration is the paradigm for particle acceleration~\citep{kotera2011astrophysics}. Diffusive shock acceleration has been proposed as the main acceleration mechanism in AGN jets~\citep{matthews2019ultrahigh} leading to a power-law energy spectrum with index around 2 and a charge-dependent exponential cutoff~\citep{supanitsky2013upper} that accounts for the limitation in the maximal energy that a source can accelerate particles
\begin{equation}
  \frac{dN}{dE dt} = L_{CR} E^{-2} \exp(-E/ZR_{cut}),
  \label{eq:spectrum}
\end{equation}
\noindent where $L_{CR}$ is the cosmic-ray luminosity, $Z$ is the particle charge, and $R_{cut}$ is the rigidity cutoff such that $E_{cut} = ZR_{cut}$. The cosmic-ray luminosity of each source and the charge of the particles are case studies of this paper, therefore in each study in this work, the values of $L_{CR}$ and $Z$ will be defined. The rigidity cutoff is related to the escape time from the source. From references~\cite{Eichmann_2018, eichmann2019high} one can relate $R_{cut}$ to the jet power $Q_{jet}$ by
\begin{equation}
  R_{cut} = 5.4 g_{ac} \sqrt{1-g_{cr}} \bigg( \frac{Q_{jet}}{10^{43}\text{ erg/s}} \bigg)^{1/2} \text{EV},
\end{equation}
\noindent where $g_{ac} = \sqrt{\frac{8\beta_{sh}^2}{f^2\beta_j}}$ depends on plasma physics details in the acceleration site, with $\beta_{sh}$ the typical shock velocity responsible for the particle acceleration (in speed of light units), $\beta_j$ the jet velocity, and $f$ provides specific plasma properties, with $1 \leq f \leq 8$ for shocks with typical geometries and turbulent magnetic fields~\citep{eichmann2019high}. Typical velocity of the shock waves $\beta_{sh} = 0.2$~\citep{matthews2019ultrahigh} is assumed for the three sources. It is estimated that $\beta_j^{\rm CenA} \sim 0.5$~\citep{hardcastle2003radio}, $\beta_j^{\rm VirA} \sim 0.9-0.99$~\citep{Hada_2013}, $\beta_j^{\rm ForA} \sim 0.4-0.9$~\citep{Maccagni_2020}. It is estimated that $Q_{jet}^{CenA} \sim 10^{43}-10^{44}$ erg/s \citep{yang2012deep,Sun_2016}, $Q_{jet}^{VirA} \sim 10^{44}-10^{45}$ erg/s \citep{Kino_2013} and, by radio flux arguments, $Q_{jet}^{ForA} \sim Q_{jet}^{VirA}$.

The factor $g_{cr}$ accounts for the relation between the energy stored in hadrons and in the magnetic field in the acceleration region~\citep{pacholczyk1970radio}, and can be written as
\begin{equation}
    g_{cr}^{-1} = 1 + \frac{3}{4}\Bigg( \frac{B}{B_{m}} \Bigg)^{7/2},
\end{equation}
\noindent where $B$ is the magnetic field strength in the acceleration region, and $B_{m}$ is the magnetic field strength in the minimum energy condition. Using $B_m^{\rm CenA} \sim 30-60\ \mu$G \citep{burns1983inner, Kraft_2002, feigelson1981x}, $B^{\rm CenA} = 23\ \mu$G~\citep{hess2020resolving} and equipartition for Virgo A \citep{Snios_2019} and Fornax A \citep{Maccagni_2020}, $B_{m}^{\rm VirA} = B^{\rm VirA}$, $B_{m}^{\rm ForA} = B^{\rm ForA}$, it is possible to calculate the maximum rigidity cutoffs $R_{cut}^{\rm CenA} = 10^{18.8}$ V, $R_{cut}^{\rm VirA} = 10^{19.3}$ V, and $R_{cut}^{\rm ForA} = 10^{19.5}$ V. As an upper-limit, the rigidity cutoff of M82 is taken as $10^{19.5}$ V.

\section{Other nuclei}
\label{app:nuclei}

\begin{figure}
  \centering
  \includegraphics[width=1.0\columnwidth]{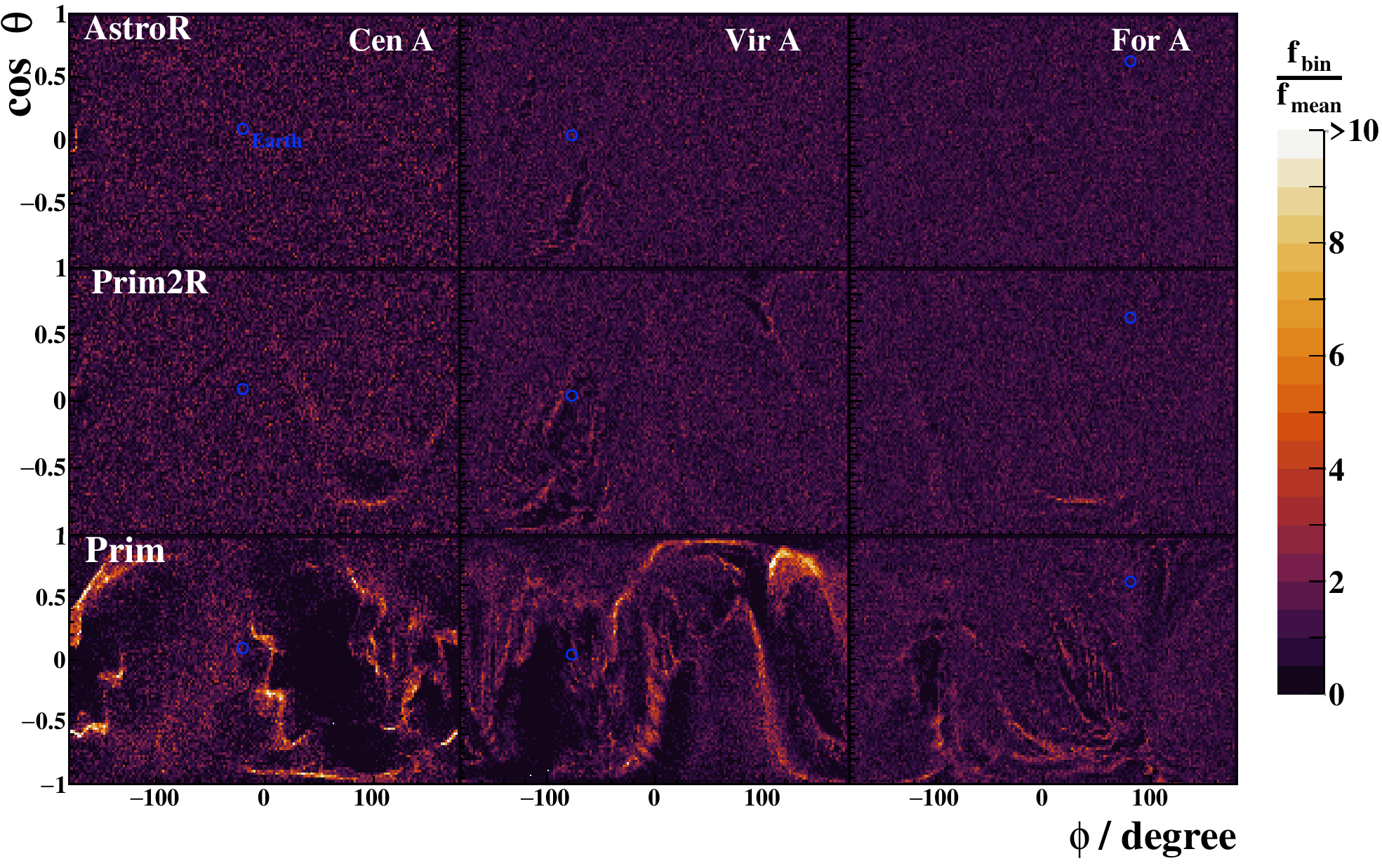}
  \caption{Relative flux of UHECR around Cen A, Vir A, and For A. Each line in the figure shows one EGMF model: AstroR, Prim2R, and Prim. Each column in the figure shows the results for one source: Cen A, Vir A, and For A. Particles were tracked from the source (center of the map) until they reached a sphere with radius equal to the distance from the source to Earth. The maps show the arrival position of all particles in this sphere. The blue circle shows the position of Earth. The color code in the maps shows the relative flux of UHECR. Only protons are considered to leave the sources with equal flux.}
  \label{fig:maps:source:center:pr}
\end{figure}

\begin{figure}
  \centering
  \includegraphics[width=1.0\columnwidth]{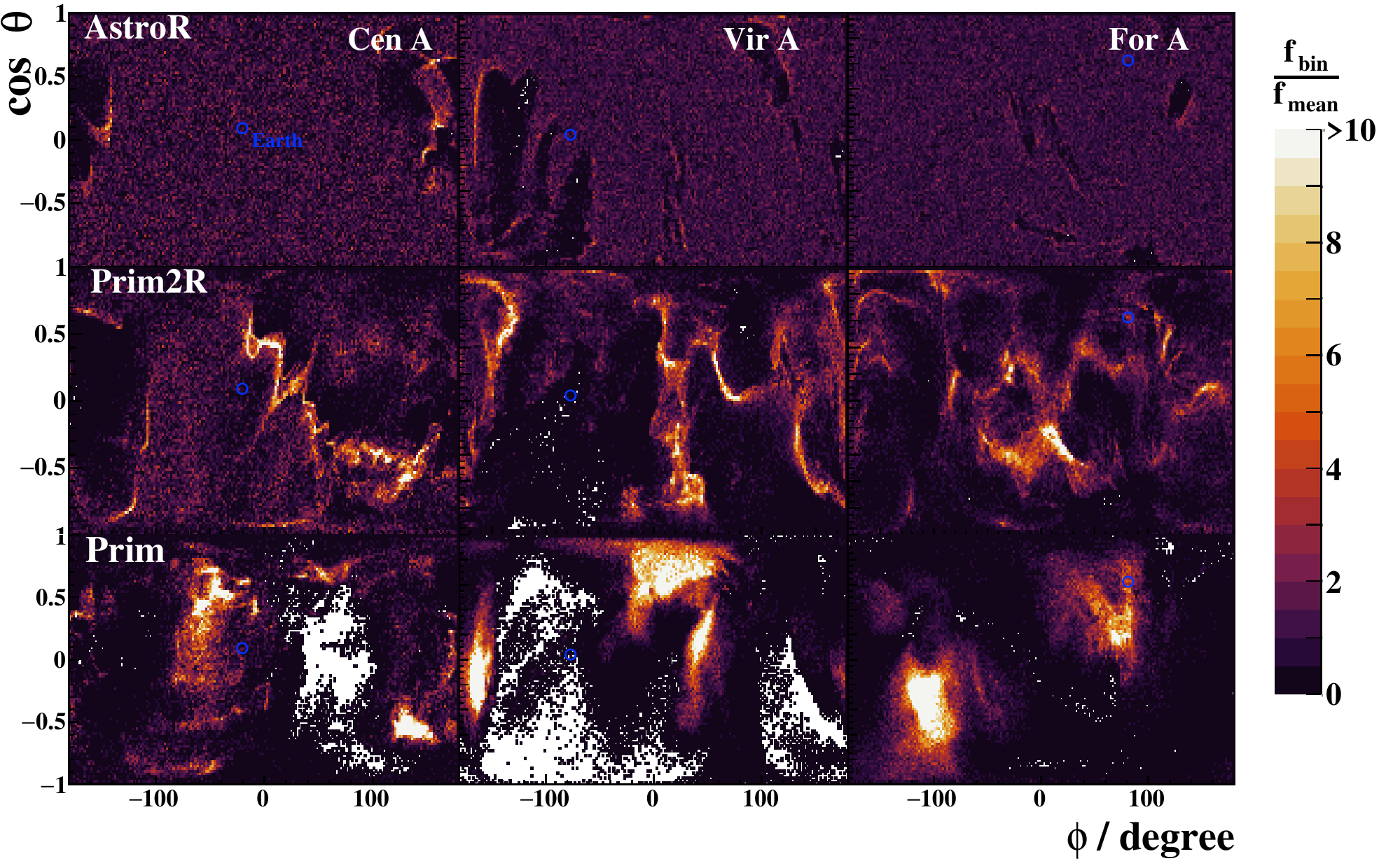}
  \caption{Relative flux of UHECR around Cen A, Vir A, and For A. Each line in the figure shows one EGMF model: AstroR, Prim2R, and Prim. Each column in the figure shows the results for one source: Cen A, Vir A, and For A. Particles were tracked from the source (center of the map) until they reached a sphere with radius equal to the distance from the source to Earth. The maps show the arrival position of all particles in this sphere. The color code in the maps shows the relative flux of UHECR. Only iron nuclei are considered to leave the sources with equal flux.}
  \label{fig:maps:source:center:fe}
\end{figure}

\begin{figure}
  \centering
  \includegraphics[trim=0 375 100 0,clip,width=1.0\columnwidth]{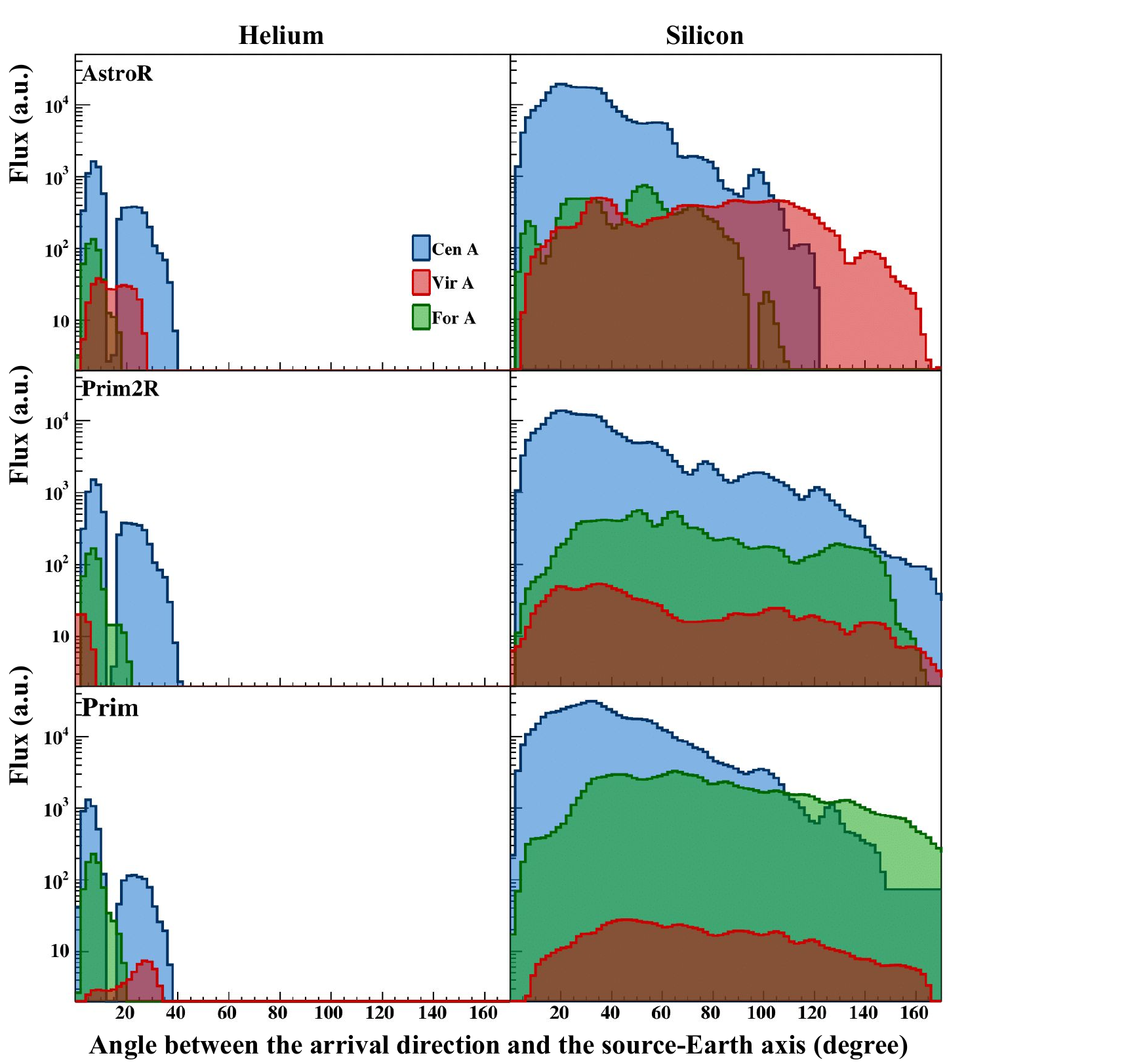}
  \caption{Angular distribution of events with $E>32$ EeV in relation to the source direction. Each line in the figure shows one EGMF model: AstroR, Prim2R, and Prim. Each column in the figure shows a different nucleus leaving the source: helium and silicon nuclei. Note that in each column of the figure, all nuclei fragments on the way to Earth are shown as arriving at Earth when only helium and silicon nuclei left the source. The three sources are shown by different colors: blue, red, and green for Cen A, Vir A, and For A. The sources are considered to output the same UHECR flux.}
  \label{fig:deflection:hist:he:si}
\end{figure}

In section~\ref{sec:arrival}, the relative flux of UHECR around each source is presented only for all nuclei with equal flux. Figures~\ref{fig:maps:source:center:pr} and~\ref{fig:maps:source:center:fe} show the same plots for proton and iron nuclei. Another representation of the enhancement/suppression effect is seen in the angular distributions of events in relation to the source direction. In section~\ref{sec:arrival}, the plots were done for proton, nitrogen and iron nuclei. Figure~\ref{fig:deflection:hist:he:si} shows the equivalent plots for helium and silicon nuclei.

\begin{figure}
  \centering
  \includegraphics[trim=0 135 0 0,clip,width=0.9\columnwidth]{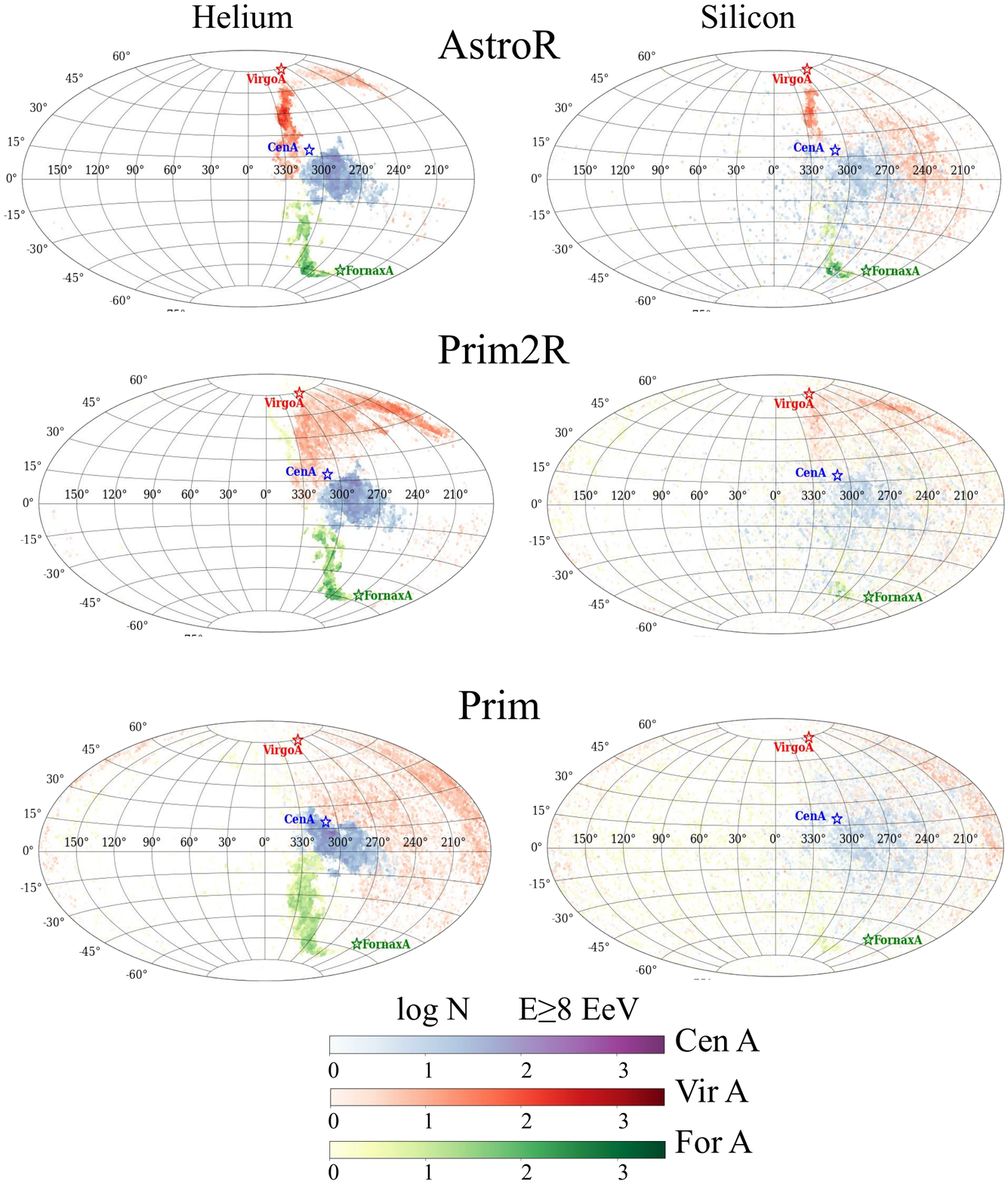}
  \caption{Sky maps in Galactic coordinates and Aitoff projections of all simulated events which arrived at Earth with energy above 8 EeV. Each line in the figure shows one of the EGMF models considered here. Each column in the figure shows a different nuclei leaving the source: helium and silicon nuclei. Note that in each column of the  figure, all nuclei fragments on the way to Earth are shown as arriving at Earth when only helium or silicon nuclei left the source. The three sources are shown as blue, red, and green stars for Cen A, Vir A, and For A, respectively. The flux of events follows the same color-code, each color representing only the events generated in the respective source.}
  \label{fig:arrival:8:he_si}
\end{figure}

\begin{figure}
  \centering
  \includegraphics[trim=0 135 0 0,clip,width=0.9\columnwidth]{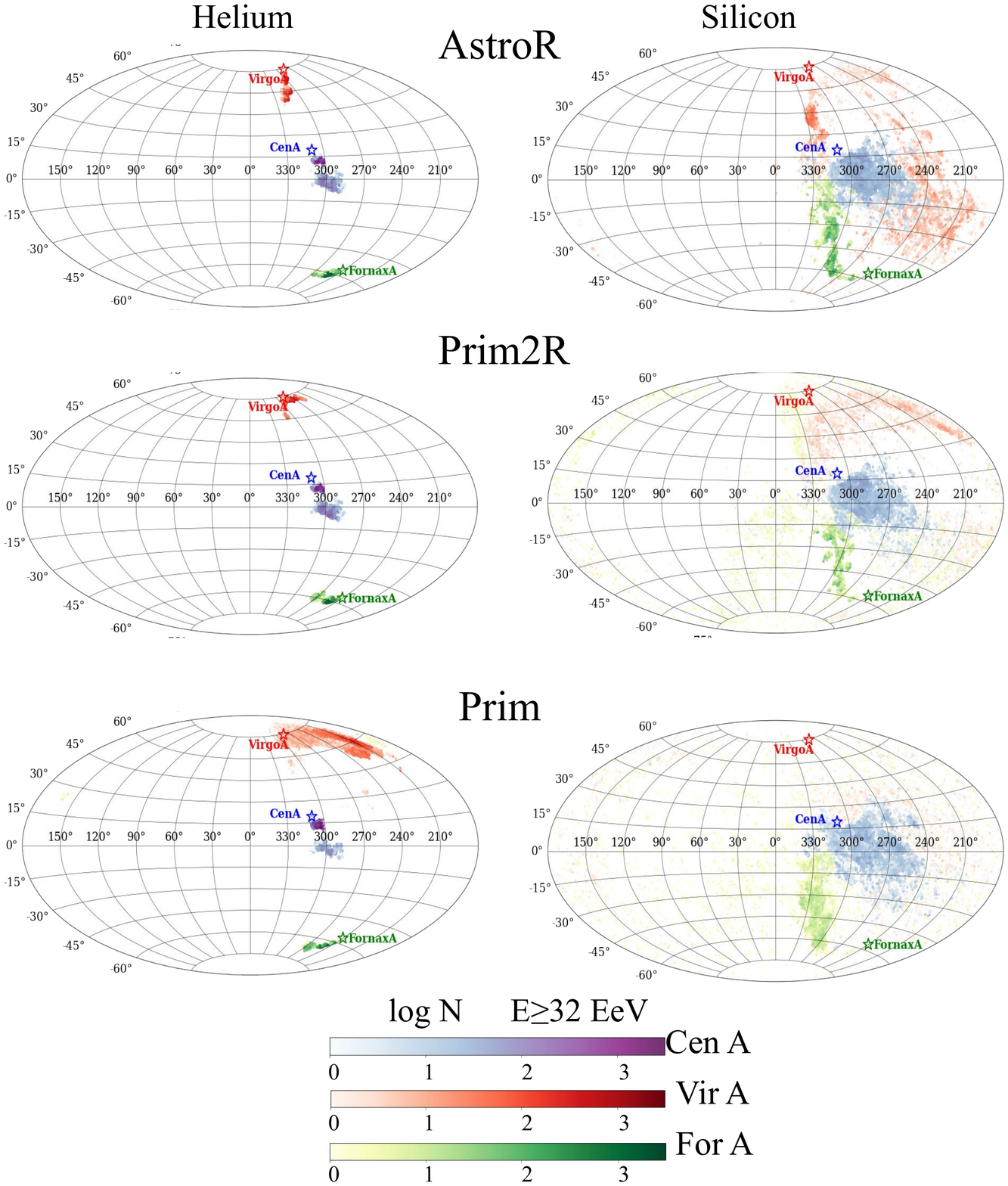}
  \caption{Sky maps in Galactic coordinates and Aitoff projections of all simulated events which arrived at Earth with energy above 32 EeV. Each line in the figure shows one of the EGMF models considered here. Each column in the figure shows a different nucleus leaving the source: helium and silicon nuclei. Note that in each column of the  figure, all nuclei fragments on the way to Earth are shown as arriving at Earth when only helium or silicon nuclei left the source. The three sources are shown as blue, red, and green stars for Cen A, Vir A, and For A, respectively. The flux of events follows the same color-code, each color representing only the events generated in the respective source.}
  \label{fig:arrival:32:he_si}
\end{figure}

\begin{figure}
  \centering
  \includegraphics[trim=0 250 50 0,clip,width=0.85\columnwidth]{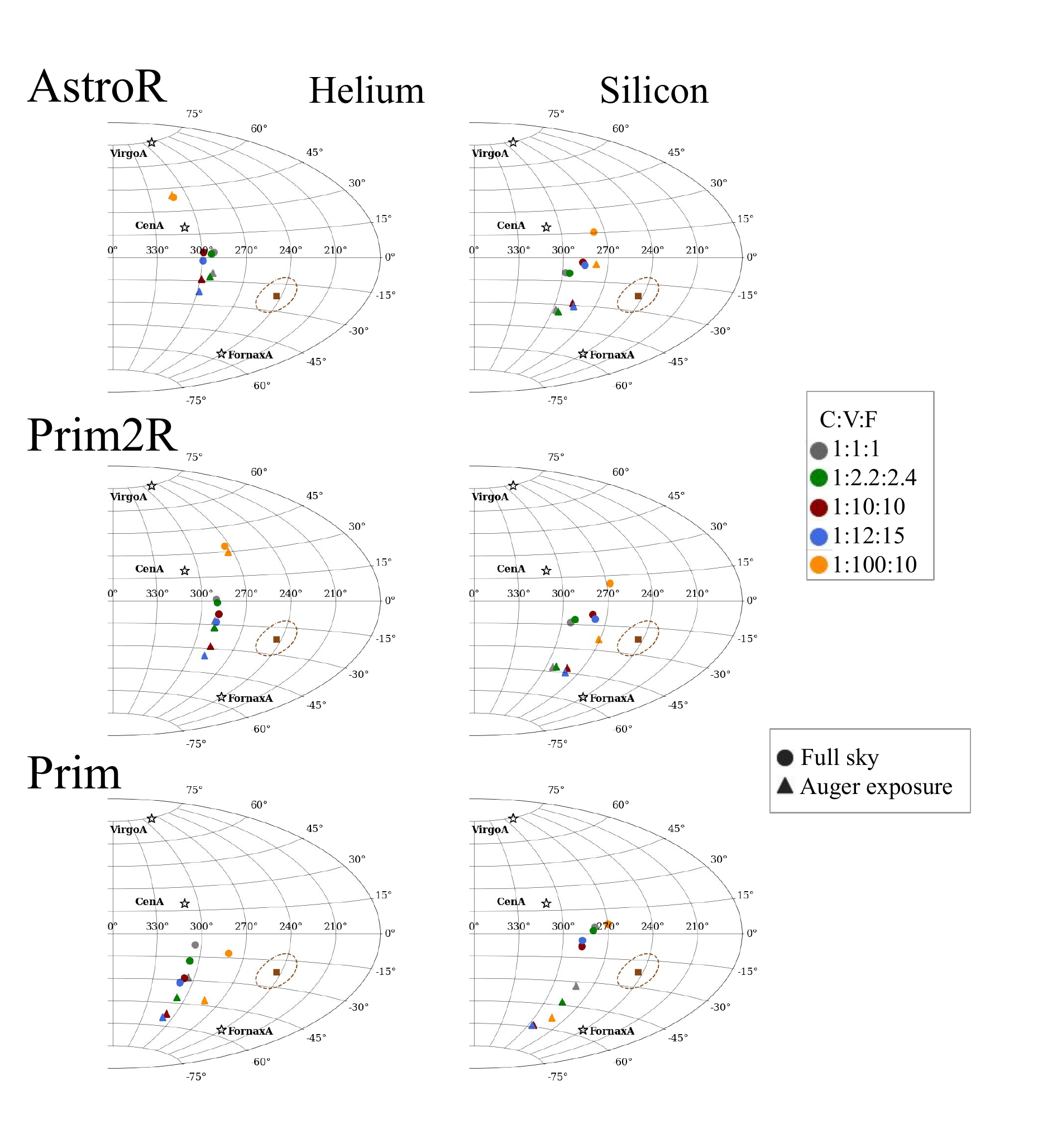}
  \caption{Sky maps in Galactic coordinates and Aitoff projections showing the dipole direction for the events with energy above 8 EeV. Each line in the figure shows one of the EGMF models considered here. Each column in the figure shows a different nuclei leaving the source: helium and silicon nuclei. Note that in each column of the  figure, all nuclei fragments on the way to Earth are shown as arriving at Earth when only helium or silicon nuclei left the source. The three sources are shown as stars. The brown square shows the direction of the dipole measured by the Pierre Auger Observatory and the dashed brown line shows its one sigma uncertainty. Colored circles show the direction of the dipoles calculated with the simulated events from Cen A, Vir A, and For A. Each color corresponds to a relation of the flux emitted by Cen A : Vir A : For A as given in the legend.}
  \label{fig:dipole:8:he_si}
\end{figure}

\begin{figure}
  \centering
  \includegraphics[trim=0 250 50 0,clip,width=0.85\columnwidth]{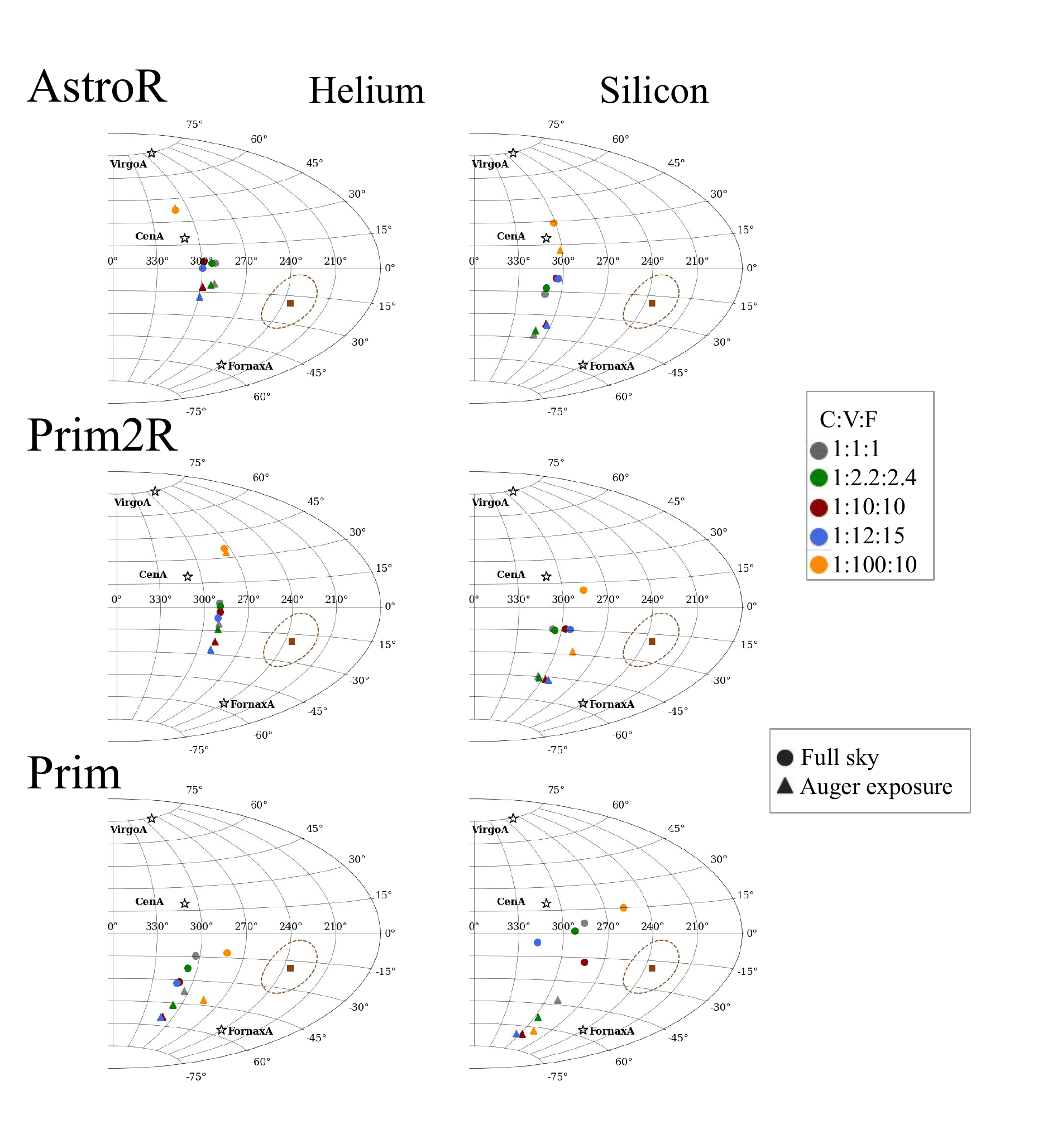}
  \caption{Sky maps in Galactic coordinates and Aitoff projections showing the dipole direction for the events with energy between 8 and 16 EeV. Each line in the figure shows one of the EGMF models considered here. Each column in the figure shows a different nuclei leaving the source: helium and silicon nuclei. Note that in each column of the  figure, all nuclei fragments on the way to Earth are shown as arriving at Earth when only helium or silicon nuclei left the source. The three sources are shown as stars. The brown square shows the direction of the dipole measured by the Pierre Auger Observatory and the dashed brown line shows its one sigma uncertainty. Colored circles show the direction of the dipoles calculated with the simulated events from Cen A, Vir A, and For A. Each color corresponds to a relation of the flux emitted by Cen A : Vir A : For A as given in the legend.}
  \label{fig:dipole:8-16:he_si}
\end{figure}

\begin{figure}
  \centering
  \includegraphics[trim=0 250 50 0,clip,width=0.85\columnwidth]{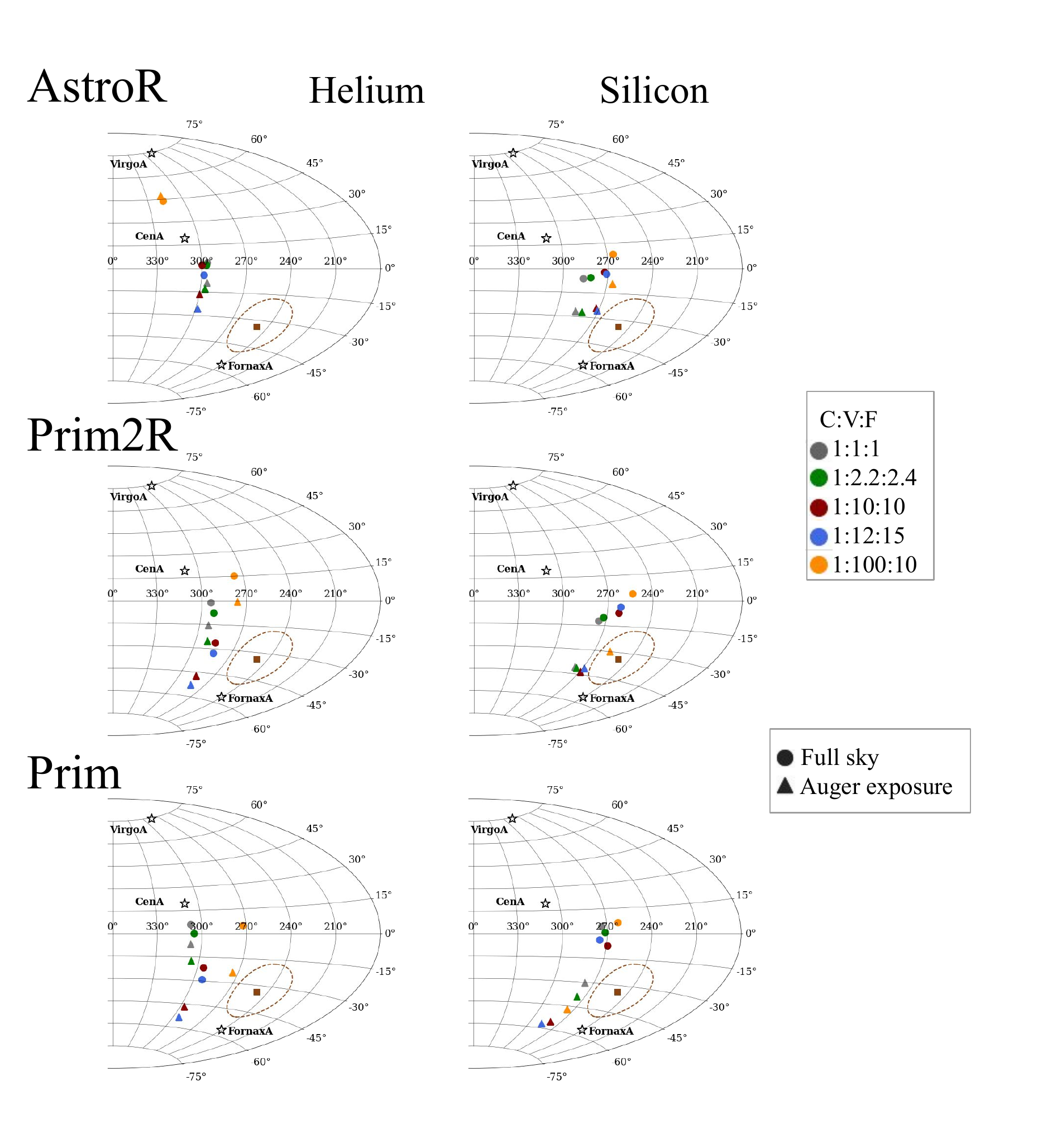}
  \caption{Sky maps in Galactic coordinates and Aitoff projections showing the dipole direction for the events with energy between 16 and 32 EeV. Each line in the figure shows one of the EGMF models considered here. Each column in the figure shows a different nuclei leaving the source: helium and silicon nuclei. Note that in each column of the  figure, all nuclei fragments on the way to Earth are shown as arriving at Earth when only helium or silicon nuclei left the source. The three sources are shown as stars. The brown square shows the direction of the dipole measured by the Pierre Auger Observatory and the dashed brown line shows its one sigma uncertainty. Colored circles show the direction of the dipoles calculated with the simulated events from Cen A, Vir A, and For A. Each color corresponds to a relation of the flux emitted by Cen A : Vir A : For A as given in the legend.}
  \label{fig:dipole:16-32:he_si}
\end{figure}

\begin{figure}
  \centering
  \includegraphics[trim=0 250 50 0,clip,width=0.85\columnwidth]{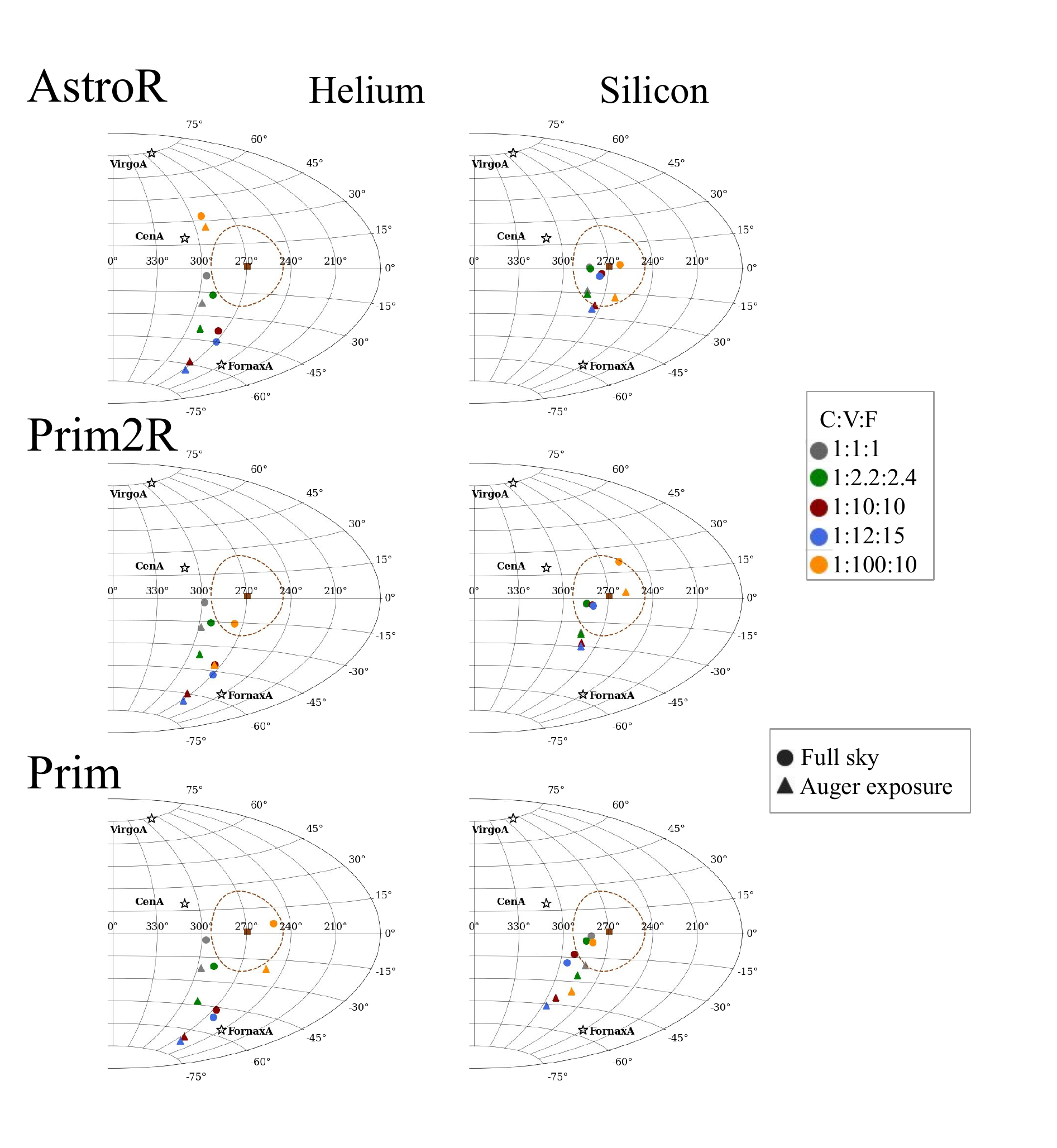}
  \caption{Sky maps in Galactic coordinates and Aitoff projections of all simulated events which arrived at Earth with energy above 32 EeV. Each line in the figure shows one of the EGMF models considered here. Each column in the figure shows a different nuclei leaving the source: helium and silicon nuclei. Note that in each column of the  figure, all nuclei fragments on the way to Earth are shown as arriving at Earth when only helium or silicon nuclei left the source. The three sources are shown as stars. The brown square shows the direction of the dipole measured by the Pierre Auger Observatory and the dashed brown line shows its one sigma uncertainty. Colored circles show the direction of the dipoles. Each color corresponds to a relation of the flux emitted by Cen A : Vir A : For A as given in the legend.}
  \label{fig:dipole:32:he_si}
\end{figure}

\begin{figure}
  \centering
  \includegraphics[trim=0 135 0 0,clip,width=0.85\columnwidth]{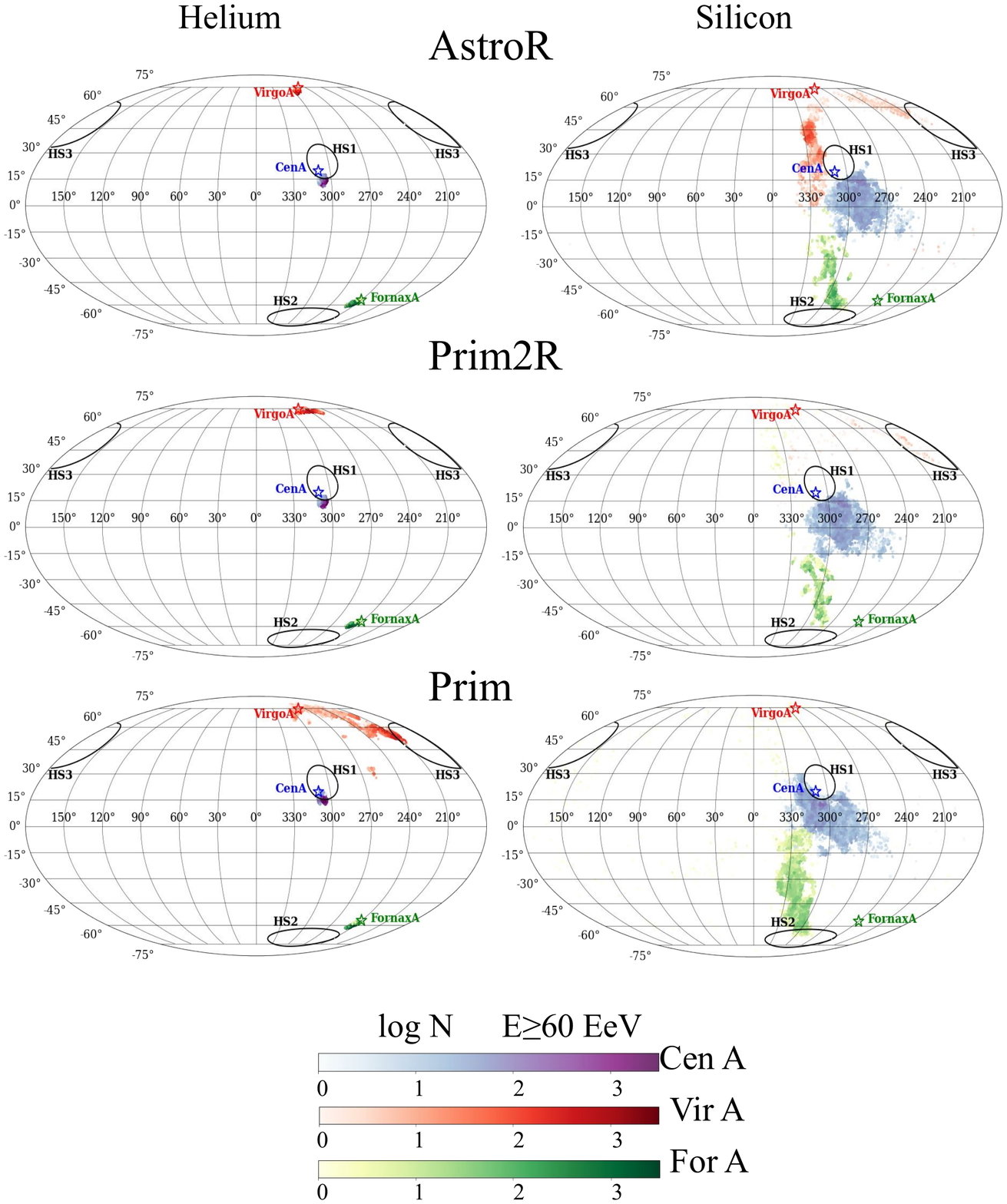}
  \caption{Sky maps in Galactic coordinates and Mollweide projections of all simulated events which arrived at Earth with energy above 60 EeV. Each line in the figure shows one of the EGMF models considered here. Each column in the figure shows a different nuclei leaving the source: helium and silicon nuclei. Note that in each column of the  figure, all nuclei fragments on the way to Earth are shown as arriving at Earth when only helium or silicon nuclei left the source. The three sources are shown as blue, red, and green stars for Cen A, Vir A, and For A, respectively. The flux of events follows the same color-code, each color representing only the events generated in the respective source. The three hotspot regions are circulated by black full lines.}
  \label{fig:arrival:60:he_si}
\end{figure}

Along sections \ref{sec:dipole} and \ref{sec:hotspots}, plots for H, N, and Fe nuclei were shown as representations of light, intermediate, and heavy compositions. He and Si nuclei were also simulated. The analysis of these simulations confirms the conclusions elaborated in the text. He nuclei plots represents the transition from light (H) to intermediate (N) nuclei. Si nuclei plots represents the transition from intermediate (N) to heavy (Fe) nuclei. If He and Si plots were shown together with those for H, N, and Fe nuclei, the figure would became very busy and the symbols sizes would be very small when compared to normal page/screen sizes. For these reasons, He and Si nuclei plots are presented here in~\cref{fig:arrival:8:he_si,fig:arrival:32:he_si,fig:dipole:8:he_si,fig:dipole:8-16:he_si,fig:dipole:16-32:he_si,fig:dipole:32:he_si,fig:arrival:60:he_si} for completeness.

\section{Dipole direction}
\label{app:dipole}
The dipole reconstruction was done from the simulated data set for each energy bin considered. From the possibles approaches to reconstruct the dipole~\citep{Ding_2021}, the method of reference~\citep{parizot_2005} was used to partial and full sky exposure. For the case of partial exposure evaluation, the established conditions by the Pierre Auger Observatory~\citep{Aab_2018} were reproduced, considering events with declination $\delta$ between $-90^\circ \leq \delta \leq 45^\circ$. The results are summarized in tables \ref{tab:8EeV}, \ref{tab:8-16EeV}, \ref{tab:16-32EeV} and \ref{tab:32EeV}.

\begin{table}[htb]
  \caption{Dipole direction in Galactic coordinates $(l,b)$ in degree. $\geq$8 EeV - partial exposure/full sky.}
  \begin{center}
  \footnotesize
    \centering
    \begin{tabular}{|c|c|c|c|c|c|c|c|c|c|} \hline
       & \multicolumn{3}{|c|}{\textbf{H}} & \multicolumn{3}{|c|}{\textbf{He}} & \multicolumn{3}{c|}{\textbf{N}} \\ \hline
      C:V:F & AstroR & Prim2R & Prim & AstroR & Prim2R & Prim & AstroR & Prim2R & Prim \\ \hline
      \multirow{2}{1.5cm}{1:1:1} & (297,-8.8) & (294,-17) & (306,-8.2) & (292,-9.6) & (290,-12) & (305,-28) & (286,-21) & (281,-32) & (278,-29) \\ \cline{2-10}
      & (297,4.2) & (294,-2.0) & (306,5.8) & (292,3.1) & (290,1.2) & (304,-7.0) & (286,-3.1) & (280,-7.9) & (276,0.6) \\ \hline

      \multirow{2}{1.5cm}{1:2.2:2.4} & (297,-14) & (291,-28) & (303,-20) & (294,-12) & (290,-16) & (307,-41) & (284,-22) & (280,-35) & (280,-34) \\ \cline{2-10}
      & (297,1.5) & (291,-9.7) & (302,-0.2) & (294,2.3) & (290,-1.2) & (307,-17) & (284,-2.4) & (278,-7.8) & (277,-0.7) \\ \hline

      \multirow{2}{1.5cm}{1:10:10} & (299,-17) & (283,-51) & (290,-48) & (299,-13) & (288,-28) & (309,-52) & (277,-23) & (274,-39) & (287,-51) \\ \cline{2-10}
       & (299,2.7) & (283,-30) & (288,-21) & (299,3.3) & (288,-8.2) & (307,-29) & (277,-1.9) & (270,-6.8) & (281,-11) \\ \hline

      \multirow{2}{1.5cm}{1:12:15} & (297,-28) & (281,-57) & (287,-55) & (299,-21) & (289,-34) & (311,-54) & (276,-24) & (275,-40) & (291,-53) \\ \cline{2-10}
      & (297,-5.6) & (281,-37) & (285,-28) & (299,-2.2) & (289,-13) & (310,-32) & (276,-2.8) & (270,-6.4) & (286,-12) \\ \hline

      \multirow{2}{1.5cm}{1:100:10} & (316,51) & (275,-7.9) & (254,0.2) & (311,41) & (274,30) & (284,-41) & (270,-4.7) & (261,-16) & (277,-40) \\ \cline{2-10}
      & (316,47) & (275,22) & (252,22) & (311,39) & (274,34) & (281,-12) & (270,12) & (258,13) & (268,6.7) \\ \hline \hline

      & \multicolumn{3}{|c|}{\textbf{Si}} & \multicolumn{3}{|c|}{\textbf{Fe}} & \multicolumn{3}{c|}{\textbf{All}} \\ \hline
      C:V:F & AstroR & Prim2R & Prim & AstroR & Prim2R & Prim & AstroR & Prim2R & Prim \\ \hline
      \multirow{2}{1.5cm}{1:1:1} & (298,-33) & (295,-42) & (283,-32) & (302,-42) & (305,-49) & (284,-43) & (294,-26) & (294,-37) & (284,-30) \\ \cline{2-10}
      & (298,-9.7) & (294,-14) & (279,4.0) & (302,-12) & (306,-15) & (278,1.2) & (294,-4.8) & (293,-8.7) & (281,2.0) \\ \hline

      \multirow{2}{1.5cm}{1:2.2:2.4} & (295,-34) & (292,-42) & (286,-43) & (298,-40) & (302,-50) & (291,-53) & (293,-25) & (290,-38) & (288,-43) \\ \cline{2-10}
      & (295,-9.9) & (291,-12) & (280,1.9) & (298,-9.4) & (302,-15) & (282,0.2) & (293,-3.2) & (289,-9.1) & (284,-4.3) \\ \hline

      \multirow{2}{1.5cm}{1:10:10} & (288,-29) & (283,-42) & (295,-59) & (289,-37) & (293,-50) & (297,-65) & (287,-26) & (283,-41) & (296,-58) \\ \cline{2-10}
      & (287,-3.0) & (280,-8.5) & (287,-8.0) & (288,-3.8) & (290,-12) & (283,-24) & (287,-1.7) & (280,-10) & (290,-19) \\ \hline

      \multirow{2}{1.5cm}{1:12:15} & (286,-30) & (282,-45) & (297,-59) & (290,-40) & (290,-51) & (301,-62) & (286,-28) & (282,-46) & (299,-60) \\ \cline{2-10}
      & (286,-4.7) & (278,-11) & (287,-4.0) & (288,-6.2) & (285,-14) & (294,-4.7) & (286,-3.4) & (279,-15) & (295,-22) \\ \hline

      \multirow{2}{1.5cm}{1:100:10} & (278,-3.7) & (271,-23) & (284,-53) & (278,-15) & (281,-40) & (292,-61) & (282,6.4) & (271,-19) & (279,-50) \\ \cline{2-10}
      & (277,16) & (268,10) & (270,5.8) & (275,18) & (275,5.5) & (274,2.0) & (281,22) & (267,14) & (268,0.5) \\ \hline \hline

      \multicolumn{10}{|c|}{Auger Experimental Data \citep{deAlmeida:20212Z}: (243,-21)} \\ \hline \hline
    \end{tabular}
    \label{tab:8EeV}
  \end{center}
\end{table}

\begin{table}[htb]
  \caption{Dipole direction in Galactic coordinates $(l,b)$ in degree. 8-16 EeV - partial exposure/full sky.}
  \begin{center}
  \footnotesize
  \centering
    \begin{tabular}{|c|c|c|c|c|c|c|c|c|c|} \hline
       & \multicolumn{3}{|c|}{\textbf{H}} & \multicolumn{3}{|c|}{\textbf{He}} & \multicolumn{3}{c|}{\textbf{N}} \\ \hline
      C:V:F & AstroR & Prim2R & Prim & AstroR & Prim2R & Prim & AstroR & Prim2R & Prim \\ \hline
      \multirow{2}{1.5cm}{1:1:1} & (297,-7.0) & (294,-12) & (308,-5.5) & (291,-9.3) & (290,-10.4) & (304,-37) & (287,-26) & (278,-39) & (277,-30) \\ \cline{2-10}
      & (297,5.0) & (294,0.9) & (308,7.0) & (291,3.3) & (290,2.4) & (303,-14) & (287,-5.6) & (277,-11) & (274,4.3) \\ \hline

      \multirow{2}{1.5cm}{1:2.2:2.4} & (298,-9.4) & (293,-20) & (306,-14) & (293,-10) & (290,-14) & (308,-46) & (283,-24) & (276,-39) & (278,-40) \\ \cline{2-10}
      & (298,4.3) & (293,-3.7) & (305,3.1) & (293,3.4) & (290,0.5) & (307,-22) & (283,-3.1) & (273,-8.2) & (273,-0.8) \\ \hline

      \multirow{2}{1.5cm}{1:10:10} & (303,-7.6) & (289,-38) & (296,-40) & (299,-12) & (289,-22) & (311,-54) & (275,-21) & (271,-38) & (285,-55) \\ \cline{2-10}
      & (303,8.1) & (289,-18) & (295,-13) & (299,4.1) & (289,-3.4) & (310,-31) & (275,0.2) & (264,-0.7) & (276,-7.6) \\ \hline

      \multirow{2}{1.5cm}{1:12:15} & (302,-17) & (288,-45) & (294,-47) & (300,-18) & (291,-27) & (312,-54) & (275,-22) & (271,-39) & (285,-55) \\ \cline{2-10}
      & (302,1.8) & (288,-24) & (292,-20) & (300,0.1) & (291,-6.9) & (312,-32) & (275,-0.7) & (263,-1.6) & (274,-9.3) \\ \hline

      \multirow{2}{1.5cm}{1:100:10} & (318,50) & (274,24) & (263,-4.8) & (310,39) & (275,34) & (285,-41) & (270,-4.8) & (261,-25) & (274,-45) \\ \cline{2-10}
      & (318,45) & (274,31) & (260,18) & (310,38) & (275,36) & (282,-12) & (270,12) & (256,9.2) & (261,7.2) \\ \hline \hline

      & \multicolumn{3}{|c|}{\textbf{Si}} & \multicolumn{3}{|c|}{\textbf{Fe}} & \multicolumn{3}{c|}{\textbf{All}} \\ \hline
      C:V:F & AstroR & Prim2R & Prim & AstroR & Prim2R & Prim & AstroR & Prim2R & Prim \\ \hline
      \multirow{2}{1.5cm}{1:1:1} & (311,-43) & (305,-44) & (291,-42) & (309,-51) & (312,-56) & (301,-64) & (297,-29) & (297,-37) & (290,-36) \\ \cline{2-10}
      & (311,-17) & (306,-14) & (286,6.7) & (311,-12) & (317,-20) & (296,-18) & (297,-5.2) & (297,-8.2) & (287,0.9) \\ \hline

      \multirow{2}{1.5cm}{1:2.2:2.4} & (310,-40) & (304,-45) & (297,-54) & (306,-50) & (312,-56) & (302,-64) & (296,-26) & (296,-40) & (292,-47) \\ \cline{2-10}
      & (311,-13) & (305,-15) & (292,1.9) & (308,-9.9) & (317,-19) & (297,-4.8) & (295,-2.1) & (295,-9.7) & (288,-4.3) \\ \hline

      \multirow{2}{1.5cm}{1:10:10} & (305,-35) & (288,-46) & (298,-65) & (301,-46) & (303,-59) & (302,-68) & (294,-24) & (287,-42) & (300,-59) \\ \cline{2-10}
      & (305,-6.1) & (297,-14) & (284,-18) & (301,3.8) & (303,-24) & (294,-47) & (293,0.9) & (284,-8.9) & (297,-17) \\ \hline

      \multirow{2}{1.5cm}{1:12:15} & (303,-36) & (295,-46) & (306,-65) & (300,-48) & (298,-61) & (302,-69) & (292,-27) & (284,-42) & (301,-63) \\ \cline{2-10}
      & (304,-6.4) & (294,-14) & (317,-5.7) & (300,-8.3) & (297,-28) & (292,-55) & (292,-1.5) & (280,-8.6) & (296,-28) \\ \hline

      \multirow{2}{1.5cm}{1:100:10} & (301,12) & (288,-28) & (289,-62) & (299,-8.9) & (292,-52) & (292,-68) & (290,14) & (274,-13) & (281,-53) \\ \cline{2-10}
      & (301,29) & (286,11) & (257,15) & (298,39) & (286,5.4) & (223,22) & (289,28) & (271,20) & (266,-0.3) \\ \hline \hline

      \multicolumn{10}{|c|}{Auger Experimental Data \citep{deAlmeida:20212Z}: (235,-19)} \\ \hline \hline
    \end{tabular}
    \label{tab:8-16EeV}
  \end{center}
\end{table}

\begin{table}[htb]
  \caption{Dipole direction in Galactic coordinates $(l,b)$ in degree. 16-32 EeV - partial exposure/full sky.}
  \begin{center}
  \footnotesize
  \centering
    \begin{tabular}{|c|c|c|c|c|c|c|c|c|c|} \hline
       & \multicolumn{3}{|c|}{\textbf{H}} & \multicolumn{3}{|c|}{\textbf{He}} & \multicolumn{3}{c|}{\textbf{N}} \\ \hline
      C:V:F & AstroR & Prim2R & Prim & AstroR & Prim2R & Prim & AstroR & Prim2R & Prim \\ \hline
       \multirow{2}{1.5cm}{1:1:1} & (297,-16) & (291,-45) & (289,-42) & (296,-8.9) & (294,-15) & (308,-6.7) & (280,-17) & (281,-30) & (279,-20) \\ \cline{2-10}
      & (297,1.4) & (291,-24) & (288,-15) & (296,3.7) & (294,-1.1) & (307,5.9) & (280,-1.2) & (281,-8.9) & (279,-1.2) \\ \hline

      \multirow{2}{1.5cm}{1:2.2:2.4} & (292,-27) & (283,-58) & (278,-57) & (297,-13) & (292,-25) & (306,-17) & (278,-20) & (280,-33) & (280,-23) \\ \cline{2-10}
      & (292,-4.6) & (283,-39) & (277,-31) & (297,2.0) & (292,-7.6) & (305,0.1) & (278,-2.5) & (280,-11) & (279,-1.8) \\ \hline

      \multirow{2}{1.5cm}{1:10:10} & (287,-36) & (271,-67) & (264,-66) & (300,-16) & (285,-47) & (297,-46) & (273,-27) & (276,-42) & (287,-39) \\ \cline{2-10}
      & (287,-9.0) & (271,-52) & (261,-44) & (300,2.0) & (285,-26) & (296,-22) & (273,-6.2) & (275,-16) & (284,-9.4) \\ \hline

      \multirow{2}{1.5cm}{1:12:15} & (284,-45) & (270,-68) & (264,-68) & (299,-26) & (283,-53) & (295,-53) & (272,-32) & (275,-47) & (289,-44) \\ \cline{2-10}
      & (284,-19) & (270,-53) & (261,-48) & (299,-4.3) & (283,-33) & (294,-29) & (272,-10) & (273,-21) & (286,-13) \\ \hline

      \multirow{2}{1.5cm}{1:100:10} & (307,60) & (274,-40) & (231,16.5) & (317,48) & (276,-0.5) & (274,-23) & (261,-11) & (255,-1.7) & (273,-28) \\ \cline{2-10}
      & (307,55) & (274,-14) & (230,30) & (317,44) & (276,15) & (272,5.0) & (261,6.9) & (254,19) & (268,4.8) \\ \hline \hline

      & \multicolumn{3}{|c|}{\textbf{Si}} & \multicolumn{3}{|c|}{\textbf{Fe}} & \multicolumn{3}{c|}{\textbf{All}} \\ \hline
      C:V:F & AstroR & Prim2R & Prim & AstroR & Prim2R & Prim & AstroR & Prim2R & Prim \\ \hline
      \multirow{2}{1.5cm}{1:1:1} & (286,-27) & (277,-41) & (277,-30) & (311,-45) & (303,-44) & (291,-38) & (292,-27) & (289,-37) & (283,-27) \\ \cline{2-10}
      & (286,-6.2) & (275,-12) & (274,4.2) & (311,-20) & (304,-13) & (286,11) & (292,-6.7) & (289,-11) & (281,2.2) \\ \hline

      \multirow{2}{1.5cm}{1:2.2:2.4} & (282,-27) & (275,-41) & (276,-39) & (309,-43) & (301,-44) & (295,-51) & (288,-27) & (285,-40) & (285,-37) \\ \cline{2-10}
      & (282,-5.8) & (272,-10) & (272,0.8) & (309,-17) & (301,-12) & (289,11) & (288,-5.8) & (284,-12) & (281,-0.3) \\ \hline

      \multirow{2}{1.5cm}{1:10:10} & (272,-24) & (270,-43) & (281,-55) & (304,-37) & (294,-40) & (300,-65) & (282,-28) & (277,-46) & (292,-54) \\ \cline{2-10}
      & (272,-2.2) & (263,-7.2) & (269,-7.3) & (304,-7.3) & (293,-5.6) & (286,3.0) & (282,-4.6) & (273,-15) & (286,-9.3) \\ \hline

      \multirow{2}{1.5cm}{1:12:15} & (271,-26) & (269,-41) & (287,-57) & (303,-38) & (290,-40) & (304,-66) & (280,-32) & (275,-47) & (293,-58) \\ \cline{2-10}
      & (271,-3.5) & (261,-3.7) & (275,-3.7) & (303,-8.3) & (288,-4.1) & (312,-7.7) & (280,-7.7) & (271,-17) & (286,-14) \\ \hline

      \multirow{2}{1.5cm}{1:100:10} & (267,-9.1) & (259,-30) & (277,-47) & (293,-8.2) & (288,-36) & (297,-63) & (274,1.9) & (265,-18) & (278,-46) \\ \cline{2-10}
      & (266,8.4) & (254,4.1) & (262,6.5) & (293,19) & (285,3.0) & (277,21) & (274,17) & (262,12) & (264,11) \\ \hline \hline

    \multicolumn{10}{|c|}{Auger Experimental Data \citep{deAlmeida:20212Z}: (248,-34)} \\ \hline \hline
    \end{tabular}
    \label{tab:16-32EeV}
  \end{center}
\end{table}

\begin{table}[htb]
  \caption{Dipole direction in Galactic coordinates $(l,b)$ in degree. $\geq$32 EeV - partial exposure/full sky.}
  \begin{center}
  \footnotesize
  \centering
    \begin{tabular}{|c|c|c|c|c|c|c|c|c|c|} \hline
       & \multicolumn{3}{|c|}{\textbf{H}} & \multicolumn{3}{|c|}{\textbf{He}} & \multicolumn{3}{c|}{\textbf{N}} \\ \hline
      C:V:F & AstroR & Prim2R & Prim & AstroR & Prim2R & Prim & AstroR & Prim2R & Prim \\ \hline
      \multirow{2}{1.5cm}{1:1:1} & (265,-63) & (264,-64) & (257,-61) & (297,-22) & (298,-18) & (297,-22) & (293,-7.6) & (292,-9.2) & (307,-23) \\ \cline{2-10}
      & (265,-45) & (264,-46) & (257,-42) & (297,-4.4) & (298,-2.6) & (297,-4.0) & (293,4.1) & (292,2.8) & (307,-5.6) \\ \hline

      \multirow{2}{1.5cm}{1:2.2:2.4} & (259,-66) & (258,-66) & (250,-64) & (291,-38) & (292,-36) & (289,-42) & (294,-7.7) & (292,-11) & (309,-37) \\ \cline{2-10}
      & (259,-49) & (258,-51) & (250,-47) & (291,-17) & (292,-15) & (289,-20) & (294,4.3) & (292,1.6) & (309,-16) \\ \hline

      \multirow{2}{1.5cm}{1:10:10} & (257,-66) & (254,-68) & (245,-65) & (275,-58) & (275,-60) & (268,-64) & (298,-8.4) & (292,-19) & (314,-52) \\ \cline{2-10}
      & (257,-50) & (254,-53) & (245,-48) & (275,-38) & (275,-41) & (267,-47) & (298,5.2) & (292,-2.9) & (313,-32) \\ \hline

      \multirow{2}{1.5cm}{1:12:15} & (255,-68) & (253,-68) & (245,-67) & (270,-63) & (270,-64) & (263,-67) & (299,-14) & (292,-24) & (315,-54) \\ \cline{2-10}
      & (255,-52) & (253,-54) & (245,-51) & (270,-45) & (270,-47) & (263,-51) & (299,2.1) & (292,-6.4) & (315,-34) \\ \hline

      \multirow{2}{1.5cm}{1:100:10} & (281,41) & (263,4.1) & (233,40) & (293,27) & (276,-41) & (252,-20) & (313,34) & (281,20) & (304,-48) \\ \cline{2-10}
      & (281,44) & (263,24) & (233,42) & (293,33) & (276,-16) & (252,5.9) & (313,34) & (281,26) & (303,-25) \\ \hline \hline

      & \multicolumn{3}{|c|}{\textbf{Si}} & \multicolumn{3}{|c|}{\textbf{Fe}} & \multicolumn{3}{c|}{\textbf{All}} \\ \hline
      C:V:F & AstroR & Prim2R & Prim & AstroR & Prim2R & Prim & AstroR & Prim2R & Prim \\ \hline
      \multirow{2}{1.5cm}{1:1:1} & (283,-13) & (285,-21) & (282,-20) & (284,-19) & (281,-32) & (277,-29) & (285,-14) & (285,-24) & (282,-24) \\ \cline{2-10}
      & (283,0.8) & (285,-3.5) & (282,-1.4) & (284,-2.2) & (280,-8.4) & (257,-0.4) & (285,0.6) & (285,-4.6) & (281,-0.6) \\ \hline

      \multirow{2}{1.5cm}{1:2.2:2.4} & (282,-16) & (285,-22) & (285,-26) & (279,-20) & (279,-35) & (280,-34) & (283,-17) & (284,-25) & (284,-32) \\ \cline{2-10}
      & (282,-0.1) & (284,-3.2) & (285,-4.5) & (279,-2.1) & (278,-8.6) & (277,-0.8) & (283,-0.3) & (283,-4.3) & (283,-4.6) \\ \hline

      \multirow{2}{1.5cm}{1:10:10} & (274,-22) & (282,-28) & (293,-41) & (270,-21) & (273,-35) & (284,-48) & (276,-22) & (278,-33) & (294,-47) \\ \cline{2-10}
      & (274,-3.3) & (281,-4.1) & (292,-13) & (270,-1.4) & (269,-2.6) & (277,-5.2) & (276,-2.6) & (276,-5.4) & (292,-15) \\ \hline

      \multirow{2}{1.5cm}{1:12:15} & (276,-24) & (281,-30) & (297,-46) & (269,-22) & (273,-37) & (291,-53) & (275,-25) & (277,-36) & (295,-51) \\ \cline{2-10}
      & (276,-4.8) & (280,-4.8) & (295,-18) & (269,-2.1) & (267,-1.9) & (285,-9.9) & (275,-4.5) & (274,-7.6) & (293,-19) \\ \hline

      \multirow{2}{1.5cm}{1:100:10} & (262,-17) & (258,4.0) & (284,-36) & (260,-18) & (260,-21) & (281,-45) & (265,-11) & (261,-6.7) & (285,-40) \\ \cline{2-10}
      & (262,2.2) & (258,21) & (280,-5.3) & (260,1.2) & (257,9.6) & (273,-0.4) & (265,6.3) & (259,16) & (281,-4.2) \\ \hline \hline

    \multicolumn{10}{|c|}{Auger Experimental Data \citep{deAlmeida:20212Z}: (269,1.4)} \\ \hline \hline
    \end{tabular}
    \label{tab:32EeV}
  \end{center}
\end{table}

\normalsize

\newpage

\end{document}